\documentclass[10pt]{IEEEtran}
\usepackage{color}
\usepackage[colorlinks, linkcolor=color1, anchorcolor=blue, citecolor=color1]{hyperref}
\IEEEoverridecommandlockouts
\usepackage{cite}
\usepackage{amsmath,amssymb,amsfonts,bm}
\usepackage{algorithmic}
\usepackage{graphicx}
\usepackage{textcomp}
\usepackage{xcolor}
\usepackage{amssymb}
\usepackage{subfigure}
\usepackage{graphicx,booktabs,multirow}
\usepackage[lined,boxed,commentsnumbered, ruled]{algorithm2e}
\usepackage{mathrsfs}
\usepackage{pgfplots}
\usepackage{mathtools}
\usepackage{float}

\definecolor{colorhkust}{RGB}{20,43,140}
\definecolor{colortsinghua}{RGB}{116,52,129}
\definecolor{color1}{RGB}{128,0,0}

\newtheorem{lemma}{Lemma}
\newtheorem{theorem}{Theorem}

\newtheorem{definition}{Definition}

\newtheorem{assumption}{Assumption}

\newcommand{\grad}{\mathrm{grad}}

\newcommand{\mini}{\operatorname{minimize}}
\newcommand{\maxi}{\operatorname{maximize}}
\newcommand{\subj}{\operatorname{subject~to}}

\newcommand{\diagg}{\operatorname{diag}}

\newcommand{\re}[1]{\textcolor{black}{#1}}
\newcommand{\changeTao}[1]{\textcolor{black}{#1}}
\date{}

\begin{document}

\title{Interference Nulling Using Reconfigurable Intelligent Surface}
\author{
\IEEEauthorblockN{Tao~Jiang, \IEEEmembership{Graduate Student Member,~IEEE} and Wei~Yu, \IEEEmembership{Fellow,~IEEE}}
\thanks{This work is supported by Huawei Technologies Canada. 
The authors are with The Edward S.~Rogers Sr.~Department of
Electrical and Computer Engineering, University of Toronto, Toronto, ON M5S 3G4, Canada
(e-mails: \{taoca.jiang@mail.utoronto.ca,  weiyu@ece.utoronto.ca\}).}}

\maketitle
%\doublespacing

\begin{abstract} 
This paper investigates the interference nulling capability of reconfigurable
intelligent surface (RIS) in a multiuser  environment where
multiple single-antenna transceivers communicate simultaneously in a shared
spectrum. From a theoretical perspective, we show that when the channels
between the RIS and the transceivers have line-of-sight and the direct paths
are blocked, it is possible to adjust the phases of the RIS elements to null
out all the interference completely and to achieve the maximum $K$
degrees-of-freedom (DoF) in the overall $K$-user interference channel, provided that the number of RIS elements exceeds some finite value that depends on $K$. Algorithmically, for any fixed channel realization we formulate the interference nulling problem as a feasibility problem, and propose an
alternating projection algorithm to efficiently solve the resulting nonconvex
problem with local convergence guarantee.  Numerical results show that the
proposed alternating projection algorithm can null all the interference if the
number of RIS elements is only slightly larger than a threshold of $2K(K-1)$.
For the practical sum-rate maximization objective, this paper proposes to use
the zero-forcing solution obtained from alternating projection as an initial
point for subsequent Riemannian conjugate gradient optimization and shows that
it has a significant performance advantage over random initializations.  For
the objective of maximizing the minimum rate, this paper proposes a subgradient
projection method which is capable of achieving excellent performance at low complexity.
\end{abstract}


\section{Introduction}
Recent emergence of reconfigured intelligent surface (RIS) for passive
beamforming has resulted in a wide range of applications for enhancing the
performance of wireless communication systems using RIS. 
This is due to the ability of the RIS to reflect incident wireless signal
toward a desired direction in a controlled manner.
In this paper, we investigate a new way of utilizing the RIS by asking
the following question: In a multiuser environment 
with multiple independent communication streams sharing the same time and
frequency resources, is it possible to configure the RIS to simultaneously 
reflect multiple beams, while completely {\em nulling interference} among them?
The main result of this paper is that this is indeed possible, provided that
the RIS has a sufficiently large number of elements.  We show that when a
multiuser interference environment with $K$ transceiver pairs is augmented 
with an RIS, the interference nulling capability of the RIS can significantly
improve the throughput for all $K$ transmissions at the same time.

When multiple transmission links share the same time and frequency resources, 
a critical challenge is to manage the interference between the links. 
From an information theoretical perspective, for a $K$-user interference channel,  
the achievable degree-of-freedom (DoF) is $K$ if there is no interference but
drops down to zero if the interference is present and is treated as noise.
Although interference alignment \cite{4567443} can recover half of
the total $K$ DoF, the alignment technique requires assumption of time-varying 
channels with symbol extension, and is difficult to implement in practice.

This paper aims to derive more practical interference nulling strategies for the
RIS-assisted interference network in that we assume constant 
channels with no symbol extensions
and derive conditions under which a full DoF of $K$ can be achieved for 
the RIS-assisted $K$-user interference channel.
Moreover, we formulate the interference nulling problem 
as a feasibility problem and propose an efficient alternating projection
method to find the RIS configuration that nulls the interference completely.
Further, we develop numerical algorithms for maximizing both the \changeTao{sum rate} and
the \changeTao{minimum rate} of the RIS-assisted interference channel. 

\re{Together, these results show the considerable promise of using the
reconfigurability of the RIS for interference mitigation in order to
accommodate a large number of users in the same time-frequency resource block.
This is an important goal for achieving
bandwidth efficient wireless multiple access, especially for
device-to-device (D2D) applications.}

\subsection{Related Works}
The RIS is typically implemented as a thin layer of electromagnetic material
composed of many passive elements that can control the phases of radio
signal reflections \cite{di2020smart}. This enables a programmable radio 
environment where the overall channel can be reconfigured by the reflection
coefficients at the RIS.  The RIS can also be thought of as a passive
beamformer, which can be used to enhance the signal-to-noise ratios (SNR) of 
wireless transmissions. For example, the RIS can be used in a multiuser MISO 
cellular network for boosting energy efficiency
\cite{huang2019reconfigurable,wu2019intelligent}. The weighted sum-rate
maximization problem is investigated in \cite{guo2020weighted} for an
RIS-assisted multiuser MISO scenario.  The capacity of the RIS-aided MIMO
communication system is considered in \cite{9110912}.  The minimum-rate
maximization problem is studied in \cite{alwazani2020intelligent,9246254,yu2020joint}. 
\re{The RIS can also be used to manipulate the artificial noises for
enhancing physical layer security as shown in \cite{yu2020robust,7572045}.}
These papers all show significant performance gains due to the deployment of 
an RIS in a wireless cellular setting in terms of various system objectives. 

This paper considers the ability of the RIS to enhance the performance of 
an interference environment with multiple independent transmitter-receiver pairs
sharing the same resource block. 
For such a $K$-user interference channel, conventional interference alignment
technique \cite{4567443} can achieve a sum-DoF of $K/2$. In the presence of
an RIS, \cite{6857425} shows that the sum-DoF can be improved from $K/2$ to $K$, if
the number of the RIS elements exceeds $K(K-1)$ and if the RIS is active, i.e.,
the RIS elements can amplify, attenuate and change the phase of the incident
signals. If the RIS is passive lossless, i.e., it can only add a phase shift to
incident signals, \cite{6857425} provides a probabilistic lower bound
for the sum DoF, which asymptotically tends to $K$ as the number of RIS
elements goes to infinity.  
These results are established for the single-antenna transceivers, under the
assumption of an i.i.d. time-varying channel model with infinite symbol extensions
(which are unrealistic to implement in practice).  For the multi-antenna transceivers,
\cite{fu2020reconfigurable}  proposes a low-rank optimization approach to
maximize the achievable DoF in a multiple-input multiple-output (MIMO)
D2D network, so that interference can be eliminated not only
by the RIS but also by the precoding and receive beamforming vectors.  In this
paper, we investigate the case where the transceivers are equipped with only 
a single antenna in
order to focus on the capability of the RIS for interference suppression.
In contrast to \cite{6857425}, we develop algorithms for fixed channel 
realizations with no symbol extensions. Further, we focus on passive 
lossless RIS, because it is more power efficient and is easier 
to implement as compared to active RIS. 

In addition to the above theoretical DoF studies, several other works have 
focused on developing algorithms for maximizing network utilities for
RIS-assisted D2D wireless communications. For example, the sum-rate
maximization problem is studied in \cite{9301375}, which considers a system
with multiple D2D pairs communicating in a single-cell network, where the RIS
is leveraged to mitigate the interference caused by D2D links. A similar
heterogeneous network is studied in \cite{ji2020reconfigurable}, which
proposes to jointly optimize the location and the phase shift of the RIS to
maximize the sum rate of the cellular and D2D networks using a deep
reinforcement learning approach. In \cite{abrardo2021mimo}, the authors propose
an iterative algorithm for optimizing the RIS to maximize the sum rate of all
the transceiver pairs by taking into account the electromagnetic properties and
the circuital implementation of the RIS. None of these works explicitly 
utilize the interference nulling capability of the RIS for network utility 
maximization. 

\re{}

\subsection{Main Contributions}

This paper investigates the interference nulling capability of the RIS and shows
that the use of an RIS
can significantly enhance the achievable throughput in a $K$-user fully
connected interference channel.  We focus on the passive lossless RIS and
show that theoretically it is possible to satisfy the interference nulling
conditions by adjusting only the phase shifts at the RIS, provided that the
number of elements at the RIS exceeds a finite value that only depends on $K$.
This theoretical result is derived under the assumption that the channels
between the RIS and the transceivers have line-of-sight and the direct paths 
between the transmitters and the receivers are blocked (which is a practical 
scenario for deploying the RIS).

From an algorithmic perspective, this paper proposes an efficient
alternating projection algorithm for finding an interference nulling (or
zero-forcing) solution for any arbitrary fixed channel realization. 
This is achieved by formulating the interference nulling condition as a feasibility problem.
Although the resulting problem is nonconvex, we show that the sequence of
solutions generated by the proposed alternating projection algorithm converges
locally. Simulation results demonstrate that the proposed alternating
projection algorithm can find a zero-forcing solution with high probability
if the number of RIS elements is slightly larger than $2K(K-1)$.

This paper also proposes practical algorithms for maximizing the network utility
in an RIS-enabled interference network. For maximizing the sum rate, we propose 
a two-stage optimization scheme, in which we first run the alternating projection algorithm to find a
solution to minimize interference, then use the resulting solution as an
initial point for subsequent Riemannian conjugate gradient (RCG) optimization, 
which can find a stationary point of the sum-rate maximization problem on the complex unit circle manifold. Simulation results show that the proposed two-stage optimization
scheme can significantly outperform all benchmark methods with random initializations.
 
Moreover, we propose a subgradient projection method to solve the minimum-rate
maximization problem. The proposed method is scalable to problems with a large 
number of RIS elements. It achieves good performance, despite having much less
complexity as compared to the previous semidefinite programming relaxation (SDR) 
or the successive convex approximation (SCA) based approaches. 

\re{
The proposed algorithms are applicable regardless of whether the channels
have line-of-sight. Further, the proposed algorithms can all be readily
extended to scenarios in which the direct paths between the transmitters and
the receivers are not blocked. 
}

\subsection{Paper Organization and Notations}

The rest of this paper is organized as follows. Section \ref{sect:system_model}
describes the system model. Section \ref{sect:IA_RIS} investigates the
interference nulling problem. Section \ref{sect:AP_IA} presents the alternating
projection algorithm. Section \ref{sect:utility} describes numerical methods 
for sum-rate and minimum-rate maximizations. Section
\ref{sec:Direct} extends the algorithms to the scenario with direct paths
between the transmitters and the receivers. Section
\ref{sect:simulations} provides simulation results. Section
\ref{sect:concllusion} concludes the paper.

The notations used in this paper are as follows. Lower-case letters are used to
denote scalars. Vectors and matrices are denoted by lower-case and upper-case
boldface letters, respectively, e.g., $\bm v\in \mathbb{C}^N$ is a
complex vector of dimension $N$, $\bm A\in\mathbb{C}^{M\times N} $ is a
$M\times N$ complex matrix.  For the matrix $\bm A$, $\bm A^\top$, $\bm A^{\sf
H}$, and $\bm A^\ast$ denote its transpose, conjugate transpose and conjugate,
respectively. We use $\Re(\cdot)$ and $\Im(\cdot)$ to denote the real and imaginary parts of the argument. We use $|v|$ to denote the magnitude of the complex scalar $v$,
and $\bm v/|\bm v|$ to denote element-wise division of the vector $\bm v$ by
its corresponding magnitude.

\section{{System Model}}\label{sect:system_model}

\begin{figure}[t]
    \centering
    \includegraphics[width=8cm]{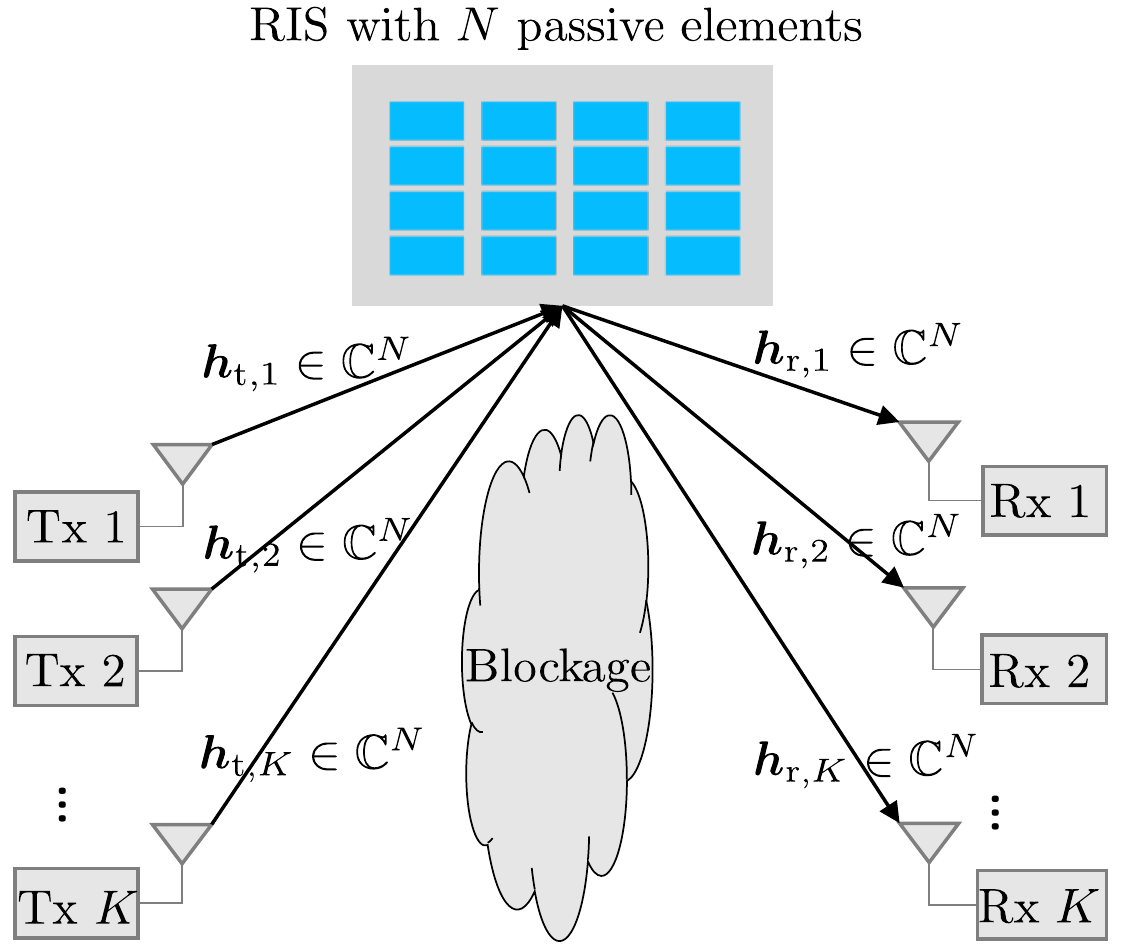}
    \caption{The RIS enabled $K$-user interference channel model.}
    \label{fig:system_model}
\end{figure}

This paper considers an RIS-enabled $K$-user interference channel, in which $K$
single-antenna transmitter and receiver pairs aim to communicate at the same
time over a common frequency band with the help of an RIS equipped with $N$ 
passive elements. We begin the
exposition by making a simplifying assumption that the direct paths are
blocked, i.e., communications have to take place through an RIS 
deployed between the transmitters and the receivers. This
assumption is relaxed in the later part of the paper where a channel model
including both the direct path and the reflection path through the RIS is considered. 

Fig.~\ref{fig:system_model} shows the channel model with a blocked direct path.
Let $\bm h_{{\rm t},j}\in\mathbb{C}^N$ denote the channel between the
transmitter $j$ and the RIS, and $\bm h_{{\rm r},k}\in\mathbb{C}^N$ denote the
channel between the RIS and the receiver $k$.  The reflection coefficients of
the RIS are denoted by $\bm v = [e^{j\mathcal{\omega}_1 },\cdots,
e^{j\omega_N}]^\top\in\mathbb{C}^N$, where $\omega_i\in (0,2\pi]$ is the phase 
shift of the $i$-th element of the RIS. 
Each transmitter $k$ sends a transmit symbol $s_k$. All 
transmissions share the same time and frequency resource. 
Each receiver $k$ receives a reflected signal through the RIS, as given by  
\begin{align}\label{eq:channel_model}
     y_k &=   \bm h_{{\rm r},k}^{\top}\diagg(\bm v) \bm h_{{\rm t},k} s_k+\sum_{j\neq k}   \bm h_{{\rm r},k}^{\top}\diagg(\bm v) \bm h_{{\rm t},j} s_j+n_k\notag\\
     &= \bm a_{k,k}^{\top}\bm v s_k+\sum_{j\neq k}  \bm a_{k,j}^{\top} \bm v s_j+n_k,
\end{align}
where 
\begin{equation}
\bm a_{k,j} \triangleq \diagg( \bm h_{{\rm t},j} ) \bm h_{{\rm r},k}
\end{equation}
is the cascaded channel from transmitter $j$ to receiver $k$, and $n_k\sim\mathcal{CN}(0,\sigma_0^2)$ is the additive Gaussian noise at receiver $k$. 
The achievable rate of the $k$-th link can now be expressed as
\begin{align}\label{eq:rate}
    R_k= \log_2\left(1+\frac{| \bm a_{k,k}^{\top}\bm v|^2 p_k}{\sum_{j\neq k} |\bm a_{k,j}^{\top}\bm v|^2 p_j+\sigma_0^2}\right),
\end{align}
where $p_k$ is the fixed transmit power level of the  $k$-th link, i.e., $\mathbb{E}[|s_k|^2]=p_k$. 

The RIS reflection coefficients can be reconfigured based on the channel state
information (CSI) for the purpose of maximizing a network utility function, e.g., 
the sum rate or the minimum rate across all the transceiver pairs. To this end,
this paper assumes that the RIS can be configured by a centralized controller that
can collect the CSI of all the links.  The CSI can be estimated through a pilot phase 
(e.g., see \cite{chen2019channel,wang2019channel,alwazani2020intelligent}). 
In this paper, we assume that the CSI is perfectly known in order to
investigate the system performance that can be achieved by optimizing the
reflection coefficients at the RIS. 

\section{Interference Nulling Capability of RIS}\label{sect:IA_RIS}

Assuming for now the channel model (\ref{eq:channel_model}) in which all the
signals go through the reflective path, we begin by investigating 
the feasibility of tuning 
the phase shifts at the RIS to completely null all the interference.
Specifically, we seek to design the RIS reflection coefficients $\bm v$ such that 
the following interference nulling conditions are satisfied:
\begin{subequations}\label{eq:zf_cond}
    \begin{align}
	    \bm a_{k,k}^{\top}\bm v \neq 0, \quad & k=1,\cdots,K, \label{eq:zf_cond00}\\
	    \bm a_{k,j}^{\top}\bm v = 0, \quad & k=1,\cdots,K, \ \ \forall j\neq k, \label{eq:zf_cond11}\\
	    |v_i| = 1, \quad & i=1,\cdots,N.\label{eq:zf_cond12}
    \end{align}
\end{subequations}
The conditions in \eqref{eq:zf_cond00} and  \eqref{eq:zf_cond11} guarantee that the powers of the desired signals are greater than zero and that all the interference signals are nulled out completely. In this case, the achievable rate of the $k$-th transceiver pair is 
%\begin{subequations}
    \begin{align}
        R_k = \log_2\left(1+\frac{| \bm a_{k,k}^{\top}\bm v|^2 p_k}{\sigma_0^2}\right), 
%            &= \log_2\frac{p_k}{\sigma_0^2}+\log_2| \bm a_{k,k}^{\top}\bm v|^2 + o\left(\log_2\frac{p_k}{\sigma_0^2}\right).
    \end{align}
%\end{subequations}
and the degree of freedom of the link $k$ is 
\begin{align}
{\rm DoF}_k =
    \lim_{\frac{p_k}{\sigma^2}\rightarrow \infty}\frac{R_k}{\log_2\frac{p_k}{\sigma_0^2}}=1.
\end{align}
Thus, the overall $K$-user interference channel achieves $K$ DoF.

Note that since the channel realizations are random, \re{we can assume that 
with high probability, none of the effective desired channels $\bm a_{k,k}$ 
are in the subspace spanned by the $K^2-K$ effective 
interfering channels $\{ \bm a_{l,j} \}_{l\neq j}$. 
Otherwise, there is no solution to the interference nulling conditions
\eqref{eq:zf_cond00}-\eqref{eq:zf_cond11}.
Since the effective channels are random vectors in $\mathbb{C}^N$, this is true
with high probability as long as $N>K^2-K$. 
}

\subsection{Feasibility of Interference Nulling}

The interference nulling is said to be feasible if there exists a vector $\bm
v$ such that the conditions in \eqref{eq:zf_cond} are met. We also call the
feasible solution a \emph{zero-forcing solution} since it nulls all the
interference. Given that the channel realizations are random, the condition
\eqref{eq:zf_cond00} typically holds with high probability. Thus, the rest of
the paper focuses on conditions \eqref{eq:zf_cond11} and \eqref{eq:zf_cond12}. 

If the complex variables $v_i$'s are treated as unconstrained variables, then
since the condition in \eqref{eq:zf_cond11} involves $K(K-1)$ complex linear
equations and $N$ complex variable $v_i$'s, we simply need $N \ge
K(K-1)$ to ensure that \eqref{eq:zf_cond11} has a solution. This would be 
the case if the RIS is active, i.e., it has both amplitude and phase control.
But, the problem with passive lossless RIS
is more complicated due to the unit modulus constraint \eqref{eq:zf_cond12}.
If we count the number of real equations and the number of real variables in 
\eqref{eq:zf_cond11} and \eqref{eq:zf_cond12}, there are $2K(K-1)$ nonlinear 
equations and $N$ real variables (i.e., $N$ phase shift $\omega_i$'s at the RIS), 
so intuitively we would need $N \ge 2K(K-1)$ to ensure the existence of a feasible
solution to \eqref{eq:zf_cond11} and \eqref{eq:zf_cond12}.

To establish the above result rigorously is however not easy, because the equations 
are nonlinear. Further, even if we establish the existence of a zero-forcing 
solution to \eqref{eq:zf_cond}, it can be challenging to actually find the
solution with low complexity. In Section~\ref{sect:AP_IA}, we propose a
computationally efficient alternating projection algorithm, which is capable 
of finding a zero-forcing solution if $N$ is slightly larger than $2K(K-1)$. 
Before presenting the alternating projection algorithm, we first provide the
following theoretical result which shows that if the channels have a line-of-sight,
then a sufficient condition for the feasibility of \eqref{eq:zf_cond} can be 
established, i.e., zero-forcing solution exists for sufficiently large $N$.

\subsection{A Sufficient Condition for Feasibility}

When the channels between the users and the RIS follow a line-of-sight model 
and the RIS elements are arranged as a rectangular array as in a typical 
implementation,
it is possible to establish a sufficient condition for zero-forcing that shows 
the conditions in \eqref{eq:zf_cond} are feasible if the number of RIS elements
$N$ exceeds some finite value that depends only on $K$.

Consider an $N_1\times N_2$ uniform rectangular array RIS with $N_1$ elements per row (horizontal direction) and $N_2$ elements per column (vertical direction). Let $\theta\in[-\frac{\pi}{2},\frac{\pi}{2}]$, $\phi\in[-\frac{\pi}{2},\frac{\pi}{2}]$ denote the azimuth angle and the elevation angle of arrival, respectively. The $n$-th element of the RIS array response vector can be written as \cite{bjornson2020rayleigh}
\begin{align}
    [\tilde{\bm a}(\theta, \phi)]_n = e^{j\frac{2\pi}{\lambda}[i_1(n)d_1\sin(\theta)\cos(\phi)+i_2(n)d_2\sin(\phi)]},
\label{eq:a_n_vector}
\end{align}
where $d_1$ and $d_2$ are the horizontal and vertical spacings, and $i_1(n) = \operatorname{mod}(n-1, N_1) $ and $i_2(n) =\lfloor (n-1)/N_2 \rfloor $ denote the horizontal index and vertical index of element $n$, respectively. The channel vector between the user $k$ and the RIS can then be written as
\begin{align}\label{eq:channel_model_ray}
    \bm h_k = \beta_k\tilde{\bm a}(\theta_k,\phi_k),
\end{align}
where $ \beta_k$ is the path-loss between the user $k$ and the RIS, and $\phi_k$ and  $\theta_k$ are the corresponding azimuth and elevation angles.

To find a sufficient condition for interference nulling, we make use of the following key lemma, which characterizes the properties of unimodular coefficient polynomials on the complex unit circle \cite{newman1990properties}.
\begin{lemma}[\cite{newman1990properties}]\label{lemma1}
    Given $z_1,\cdots,z_n$ on the complex unit circle $\mathcal{C}=\{x\in\mathbb{C}: |x|=1 \} $,  there exists a polynomial $f$ of degree $\sum_{i=1}^{n} 4^{i-1}$ with unit modulus coefficients  such that the points $z_1,\cdots,z_n$  are the only  zeros of $f$ on $\mathcal{C}$.
\end{lemma}

We establish the following sufficient condition for interference nulling for the line-of-sight channel model.
\begin{theorem}\label{proposition:feasibility_condition}
    For an $N_1\times N_2$ rectangular RIS, assuming that the channels between the RIS and the users are given by \eqref{eq:channel_model_ray}, there exists a feasible solution to the interference nulling conditions in \eqref{eq:zf_cond} if $\min(N_1, N_2)\ge \sum_{k=1}^{K(K-1)} 4^{k-1}$.
\end{theorem}
\begin{IEEEproof}
We begin with the case $N_1=1$, which corresponds to the uniform linear array case. In this case, the channel between the RIS and transmitter $j$ can be written as
\begin{align}
    \bm h_{{\rm t},j} & =\beta_j [1,\cdots, e^{j{\frac{2\pi d_2(N_2-1)}{\lambda}}\sin(\phi_{{\rm t},j})}]^\top,
\end{align}
and the channel between the RIS and receiver $k$ can be written as
\begin{align}
    \bm h_{{\rm r},k}  & = \beta_k [1,\cdots, e^{j{\frac{2\pi d_2 (N_2-1)}{\lambda}}\sin(\phi_{{\rm r},k})}]^\top.
\end{align}
Thus, we have 
\begin{subequations}
    \begin{align}
        \bm a_{k,j} &= \beta_k\beta_j [1, \cdots,e^{j{\frac{2\pi d_2 (N_2-1)}{\lambda}}(\sin(\phi_{{\rm t},j})+\sin(\phi_{{\rm r},k}))}]^\top\\
        &\triangleq \beta_k\beta_j [1, z_{k,j}, z_{k,j}^2, \cdots, z_{k,j}^{N_2-1}]^\top,
    \end{align}
\end{subequations}
where $z_{k,j} = e^{j{\frac{2\pi d_2 }{\lambda}}(\sin(\phi_{{\rm t},j})+\sin(\phi_{{\rm r},k}))}$.

Let $f(z)$ be a polynomial of degree $N_2$ with unit modulus coefficients as follows:
\begin{align}\label{eq:poly}
    f(z)=v_1+v_2z+v_3z^2+\cdots+v_{N_2}z^{N_2-1}, 
\end{align}
where $|v_i| = 1,~ \forall i$. 
The interference nulling conditions in \eqref{eq:zf_cond00}, \eqref{eq:zf_cond11},  and \eqref{eq:zf_cond12} become
\begin{subequations}\label{eq:poly_ia_cond}
    \begin{align}
        &f(z_{k,k}) \neq 0, \quad k=1,\cdots,K, \\
        &f(z_{k,j}) = 0, \quad k=1,\dots,K, \ \ \forall j\neq k, 
    \end{align}
\end{subequations}
which transform the original interference nulling problem to the problem of finding a polynomial as in \eqref{eq:poly} such that $z_{k,j}$'s are the roots, while the polynomial does not vanish at $z_{k,k}$'s. By Lemma \ref{lemma1}, we know such a polynomial exists if $N_2 = \sum_{k=1}^{K(K-1)} 4^{k-1}$.

To generalize the above result to the rectangular array case with $N_1>1$, we make the argument that a rectangular uniform array can be viewed as $N_1$ columns of the uniformly linear array of size $N_2\times 1$. Thus, if we set all the columns of the RIS to have the same reflection coefficients that null all the interference, which can be achieved if $N_2 = \sum_{k=1}^{K(K-1)} 4^{k-1}$, the entire rectangular array would also achieve the interference nulling condition. 
Mathematically, this means that the same reflection coefficients are used $N_1$ times to produce a zero-forcing solution for all $N_1$ sub-vectors of $\tilde{\bm a}(\theta, \phi)$ 
as in \eqref{eq:a_n_vector}.

Finally, if we exchange the roles of $N_1$ and $N_2$, it is easily seen that the interference nulling conditions are also achievable if $N_1 = \sum_{k=1}^{K(K-1)} 4^{k-1}$. This completes the proof. 
\end{IEEEproof}

%\begin{remark}
In Theorem~\ref{proposition:feasibility_condition}, the number of RIS elements $N$ needs to scale exponentially with the number of transceiver pairs $K$ in order to ensure feasibility for zero-forcing. But, this is only a sufficient condition. 
In the next section, we develop practical algorithms that can find feasible solutions with high probability if the number of RIS elements $N$ is only slightly larger than $2K(K-1)$, which is much smaller than the value required in Theorem~\ref{proposition:feasibility_condition}. 

%\end{remark}
\re{ %\begin{remark}
    We remark that the line-of-sight assumption on the channel model is made in Theorem \ref{proposition:feasibility_condition} in order to prove the mathematical result. The algorithmic developments in the rest of the paper do not assume the line-of-sight channel model. Further, the algorithms can be extended to the case where the direct paths are not blocked. 
 %\end{remark}
 }

\section{Alternating Projection Algorithm for Interference Nulling}\label{sect:AP_IA}

We now focus on the numerical algorithm for finding a zero-forcing solution $\bm v$
in (\ref{eq:zf_cond}). Specifically, we focus on the interference nulling condition
and formulate the following feasibility problem:
\begin{subequations}\label{prob:IA_feasibility_0}
    \begin{align}
        &~~~\operatorname{find}\quad && \bm v \\
        &\subj\quad &&\bm a_{k,j}^{\top}\bm v = 0, \ k=1,\cdots,K, \ \forall j\neq k, \\
        &\quad&&|v_i| = 1,~i=1,\cdots,N.
    \end{align}
\end{subequations} 
For ease of presentation, we define the matrices $\bm A_k \in\mathbb{C}^{N\times(K-1)}$ and  $\bm A \in\mathbb{C}^{N\times(K-1)K}$ as:
\begin{subequations}
    \begin{align}
        &\bm  A_k = [\bm a_{k,1},\hdots,\bm a_{k,k-1},\bm a_{k,k+1},\hdots,\bm a_{k,K}],\\
        &\bm A = [\bm A_1,\cdots,\bm A_K],
    \end{align}
\end{subequations}
where $\bm A_k$ contains all the interference channels to the $k$-th receiver and $\bm A$ contains all the interference channels of all the $K$ transceiver pairs. 

Then, \eqref{prob:IA_feasibility_0} can be rewritten more compactly as
\begin{subequations}\label{prob:IA_feasibility_1}
        \begin{align}
            &~~~\operatorname{find}\quad && \bm v \\
            &\subj\quad &&\bm  A^\top \bm v = \bm  0, \\
            &\quad&&|v_i|= 1,~i=1,\cdots,N.
        \end{align}
\end{subequations} 
This is a nonconvex optimization problem due to the unit modulus constraints, 
and is in general not easy to solve. In the following, we propose an efficient 
alternating projection method for solving (\ref{prob:IA_feasibility_1}).

\subsection{Alternating Projection Method}

We begin by defining the following two constraint sets
\begin{subequations}
    \begin{align}
        \mathcal{S}_1 &= \{\bm v: \bm  A^\top \bm v = \bm  0\}, \label{eq:omega_1}\\
        \mathcal{S}_2 &= \{\bm v: |v_i| = 1, ~i=1,\cdots,N\}\label{eq:omega_2},
    \end{align}
\end{subequations}
and rewrite problem \eqref{prob:IA_feasibility_1} equivalently as 
\begin{equation}\label{prob:IA_feasibility_2}
    \begin{aligned}
        &~~~\operatorname{find}\quad && \bm v \\
        &\subj\quad &&\bm v\in\mathcal{S}_1\cap\mathcal{S}_2. \\
    \end{aligned}
\end{equation} 
To find a point in the intersection of $\mathcal{S}_1$ and $\mathcal{S}_2$, the alternating projection algorithm starts with an initial $\bm v^0$, then alternatively projects onto $\mathcal{S}_1$ and $\mathcal{S}_2$ as follows:
\begin{subequations}
    \begin{align}
        \tilde{\bm v}^{t} &= \Pi_{\mathcal{S}_1}(\bm v^t),\label{eq:alter_sequence1}\\
        \bm v^{t+1} &= \Pi_{\mathcal{S}_2}(\tilde{\bm v}^{t}).\label{eq:alter_sequence2}
    \end{align}
\end{subequations} 
Here, the projection operation $\Pi_{\mathcal{S}} (\bm v)$ is defined as the point in the set $\mathcal{S}$ that has the minimum Euclidean distance to $\bm v$, which can be obtained by solving the following problem
\begin{equation}\label{prob:projection}
    \begin{aligned}
        &\underset{\bm x}{\mini}\quad && \|\bm v-\bm x \|_2^2\\
        &\subj\quad && \bm x\in\mathcal{S}.\\
    \end{aligned}
\end{equation}
Fortunately, the projections to the sets $\mathcal{S}_1$ and $\mathcal{S}_2$  have simple analytical expressions, as given by
\begin{subequations}
    \begin{align}
        \Pi_{\mathcal{S}_1}(\bm v) &= \bm v - \bm A^\ast (\bm A^\top \bm A^{\ast})^{-1}\bm A^\top \bm v, \label{eq:proj_s1} \\
        \Pi_{\mathcal{S}_2}(\bm v) &= \bm v/|\bm v|, \label{eq:proj_s2}
    \end{align} 
\end{subequations}
In \eqref{eq:proj_s1} the channel matrix $\bm A$ is assumed to be full \changeTao{column} rank
(otherwise the matrix $\bm A$ can be replaced by a new matrix $\bm A^\prime$
constructed from the basis of the \changeTao{column} space of $\bm A$). In \eqref{eq:proj_s2},
if some elements of the vector $\bm v$ are zero, these elements can be
projected to any random point on the complex unit circle.  The alternating
projection algorithm for solving problem \eqref{prob:IA_feasibility_2} is
summarized in Algorithm \ref{algo:alter}.

\begin{algorithm}[t]
    \caption{Alternating Projection for Solving \eqref{prob:IA_feasibility_1}}\label{algo:alter}
    \KwIn{Initial point $\bm v\in\mathbb{C}^N$. Channel matrix $\bm A$.}
    Initialization: $\bm v^0 = \bm v$\\
    \For{$t=0, 1,2,\cdots$}{
        $\tilde{\bm v}^{t} = \Pi_{\mathcal{S}_1}(\bm v^t)$\\
        $\bm v^{t+1} = \Pi_{\mathcal{S}_2}(\tilde{\bm v}^{t})$\\
        \If{stopping criterion is satisfied}{break}
    }
    \KwOut{$\bm v^{t+1}$}
\end{algorithm}

\subsection{Initialization}\label{subsection:initialization}

A simple way to initialize the alternating projection algorithm
is to generate a random unit modulus vector $\bm v$, with i.i.d. uniform random phases 
in $(-\pi,\pi]$.  However, the alternating projection algorithm only
nulls the interference among all the transceiver pairs; it makes no attempt
to maximize the received powers between the intended transceiver pairs. We point out
here that since it is desirable to seek a zero-forcing solution that can also
maximize the receive powers of the intended signals, instead of random initialization, 
the following initialization point could lead to better performance.

The sum power of the intended signals across all the links can be written as
\begin{align}\label{eq:direct_power}
   \sum_{k=1}^K | \bm a_{k,k}^{\top}\bm v|^2 = \bm v^{\sf H} \bm R \bm v,
\end{align}
where $\bm R = \sum_{k=1}^K \bm a_{k,k}^{\ast}\bm a_{k,k}^{\top}$, and $\bm v$
is subject to the unit modulus constraints. If we relax the unit modulus
constraints to $\|\bm v\|^2_2 = N$, then the sum of  intended signal powers
\eqref{eq:direct_power} is maximized by the eigenvector $\bm v_{\rm max}$
associated with the largest eigenvalue of the matrix $\bm R$. Experimentally, 
using $\bm v_{\rm max}$ as an initial point, followed by a projection to the
unit modulus constraint set $\mathcal{S}_2$ by \eqref{eq:proj_s2}, is found to
achieve higher sum rate as compared to random initialization. This is shown in
the numerical simulation results in Section~\ref{sect:simulations}.

\subsection{Convergence Analysis}

The convergence of the alternating projection algorithm would have been easy
to establish if $\mathcal{S}_1$ and $\mathcal{S}_2$ were convex sets. But in 
our setting, $\mathcal{S}_2$ is the complex unit circle, which is not
convex.  Further, the projection operation onto $\mathcal{S}_2$ may not be
unique (the non-uniqueness occurs when $\bm v$ has a zero component). So in general,
even the local convergence of the alternating projection algorithm is not
trivial to establish. This section presents a convergence analysis of the
proposed algorithm, which is based on the following mathematical definitions.

\begin{definition}[Semialgebraic set]
    A set $\Omega\subset\mathbb{R}^N$ is a semialgebraic set if  there exists a finite
    number of real polynomial functions $\varrho_{ij} $ and $\varsigma_{ij}$ such that 
    \begin{align}\label{eq:semialgebraic}
        \Omega = \bigcup_j \bigcap_i\{\bm x\in\mathbb{R}^N: \varrho_{ij}(\bm x) = 0, \varsigma_{ij}(\bm x)<0 \}. 
    \end{align}
\end{definition}

\begin{definition}[Prox-regular \cite{lewis2009local}]
    A set $\Omega$ is prox-regular at a point $\bar{\bm x}\in\Omega$ if the projection mapping $P_\Omega$ is single-valued around $\bar{\bm x}\in\Omega $.
\end{definition}

The semialgebraic property of the sets $\mathcal{S}_1, \mathcal{S}_2$ gives nice intersection geometry between the two sets, which is called \emph{separable intersection}  in \cite{noll2016local}.  Prox-regular property gives nice local geometry properties for the individual set \cite{noll2016local}.  The geometry of the intersection of the two sets and the regularity of individual sets play a critical role in deriving local convergence guarantee for the alternating projection algorithm. We refer to \cite{kruger2018set} for a detailed discussion on the local convergence theorem of the alternating projection algorithm. The following local convergence theorem for general semialgebraic sets is taken from \cite{noll2016local}.

\begin{lemma}[Local convergence for semialgebraic sets {\cite[Corollary 8 and Corollary 3]{noll2016local}}]\label{lemma_converge}
    Let $\mathcal{S}_1, \mathcal{S}_2$ be semialgebraic sets. Suppose $\mathcal{S}_2$ is prox-regular. Then there exists a neighborhood $\mathcal U$ of $\bm x^\ast \in\mathcal{S}_1\cap \mathcal{S}_2 $ such that every alternating projection sequence given by \eqref{eq:alter_sequence1} and \eqref{eq:alter_sequence2} which enters $\mathcal U$ converges to some point in $\mathcal{S}_1\cap \mathcal{S}_2$ with rate $\mathcal{O}(k^{-\rho})$ for some $\rho\in(0,\infty)$.
\end{lemma}

We now state the local convergence theorem for the proposed alternating projection method in Algorithm~\ref{algo:alter}.

\begin{theorem}\label{theorem:ap_convergence}
    Let $\mathcal{S}_1$ and $\mathcal{S}_2$ be defined by \eqref{eq:omega_1} and \eqref{eq:omega_2} with nonemtpy intersections, i.e.,  $\exists \bar{\bm x}\in\mathcal{S}_1\cap\mathcal{S}_2 $. With an initial point sufficiently close to $\bar{\bm x}$, the alternating projection sequence generated by \eqref{eq:alter_sequence1} and \eqref{eq:alter_sequence2} is guaranteed to converge to a point in $\mathcal{S}_1\cap\mathcal{S}_2 $ with rate $\mathcal{O}(k^{-\rho})$ for some $\rho\in(0,\infty)$.
\end{theorem}
\begin{IEEEproof}
    The set  $\mathcal{S}_1$ can be equivalently written as 
    \begin{align}
        \mathcal{S}_1 = \{\changeTao{\bm x}\in\mathbb{R}^{2N}: \tilde{\bm A}\bm x = \bm 0 \},
    \end{align} 
    where 
    \begin{align}
        \tilde{\bm A} = \begin{bmatrix}
            \Re{\bm A^\top} &-\Im{\bm A^\top}\\
            \Im{\bm A^\top} &\Re{\bm A^\top}
        \end{bmatrix},
    \end{align}
    which is a representation of the form \eqref{eq:semialgebraic}. Hence, $\mathcal{S}_1$ is a semialgebraic set. The set $\mathcal{S}_2$ can be equivalently written as 
    \begin{align}
        \mathcal{S}_2 = \bigcap_{i=1}^{N}\{ [\bm x,\bm y]^\top\in\mathbb{R}^{2\times N}: x_i^2+y_i^2 - 1=0  \}.
    \end{align}
    Hence, $\mathcal{S}_2$ is also a semialgebraic set. Since $\mathcal{S}_1$ is a convex set,  the projection onto  $\mathcal{S}_1$ from any point is single valued. $\mathcal{S}_1$ is therefore prox-regular at the point $\bar{\bm x}$. The proof then follows from Lemma~\ref{lemma_converge}.
\end{IEEEproof}

\subsection{Discussions on Global Convergence Guarantee}

Theorem \ref{theorem:ap_convergence} establishes a local convergence guarantee for
the alternating projection algorithm. 
In general, it is difficult to establish global convergence guarantee from any initial starting point for alternating projection  to nonconvex sets. 
However, it is observed in simulations that the proposed alternating projection algorithm always converges linearly to a feasible zero-forcing solution from any random initial point. To provide some intuitions about the convergence behavior, we discuss some conditions to guarantee the convergence of the alternating projection algorithm for general nonconvex sets. Specifically, the convergence of alternating projection algorithm for nonconvex sets can be established if  $\mathcal{S}_1$ and $\mathcal{S}_2$ satisfy the following assumption \cite[Theorem 1]{zhu2018convergence}.

\begin{assumption}[\cite{zhu2018convergence}]\label{assumption:nonconvex_alter}
    Let $\mathcal{S}_1$ and $\mathcal{S}_2$ be any two closed semialgebraic sets, and let $\{(\tilde{\bm v}^t,\bm v^{t+1} )\}$ be the sequence of iterates generated by the alternating projection method. Assume that the sequence $\{(\tilde{\bm v}^t,\bm v^{t+1} )\}$ is bounded and there exist subsets $\bar{\mathcal{S}}_1\subset \mathcal{S}_1$ and $\bar{\mathcal{S}}_2\subset \mathcal{S}_2$ and $t_0\in\mathbb{N} $ such that $\tilde{\bm v}^t\in\bar{\mathcal{S}}_1$ and $\bm v^{t+1}\in\bar{\mathcal{S}}_2$ for all $t\ge t_0$. Furthermore, we assume that the sets ${\mathcal{S}}_1$, ${\mathcal{S}}_2$ and subsets $\bar{\mathcal{S}}_1$, $\bar{\mathcal{S}}_2$ obey the following properties:
    \begin{itemize}
        \item Three-point property: There exists a nonnegative function $\delta_{\alpha}: \mathcal{S}_2\times \mathcal{S}_2 \rightarrow \mathbb{R}$ with $\alpha>0$ such that for all $\bm v,\bm v^\prime\in\mathcal{S}_2$ we have $ \delta_{\alpha}(\bm v,\bm v^\prime)\ge \alpha \|\bm v-\bm v^\prime \|_2^2 $ and for all $\bm v\in\bar{\mathcal{S}_2}$, $\tilde{\bm v}\in\mathcal{S}_1, \bm v^\prime = \Pi_{\mathcal{S}_2}(\tilde{\bm v}) $, we have
        \begin{align}\label{eq:3point}
            \delta_{\alpha}(\bm v,\bm v^\prime)+\|\tilde{\bm v}-\bm v^\prime \|_2^2\le \|\tilde{\bm v}-\bm v \|_2^2,
        \end{align}
        \item Local contraction property of $\mathcal{S}_1$ with respect to $\mathcal{S}_2$: There exist $\epsilon>0$ and $\beta>0$ such that 
        \begin{align}
            \|\Pi_{\mathcal{S}_1}(\bm v) - \Pi_{\mathcal{S}_1}(\bm v^\prime) \|_2\le \beta\|\bm v-\bm v^\prime \|_2,  
        \end{align}
        for all $\bm v,\bm v^\prime\in\mathcal{S}_2, \|\bm v-\bm v^\prime\|_2\le \epsilon$.
    \end{itemize}
\end{assumption}
If Assumption~\ref{assumption:nonconvex_alter} holds, the sequence generated by the alternating projection algorithm can be shown to converge to a critical point of the following  problem \cite{zhu2018convergence}: 
\begin{align}\label{prob:global_conv}
    \underset{\bm v,\tilde{\bm v}}{\min}~~ \|\bm v-\tilde{\bm v} \|_2^2+\mathbb{I}_{\mathcal{S}_1}(\bm v)+\mathbb{I}_{\mathcal{S}_2}(\tilde{\bm v}),
\end{align}
where $\mathbb{I}_{\mathcal{S}}(\bm v)= 0$ if $\bm v\in\mathcal{S}$, 
and otherwise $\mathbb{I}_{\mathcal{S}}(\bm v)= \infty$.
This is an equivalent formulation of the problem \eqref{prob:IA_feasibility_2}.

It is clear that the local contraction property of $\mathcal{S}_1$ with respect to $\mathcal{S}_2$ always holds with $\beta=1$  since $\mathcal{S}_1$ is a convex set. With respect to the three-point property, let $\bm v\in{\mathcal{S}_2}$, $\tilde{\bm v}\in\mathcal{S}_1, \bm v^\prime = \Pi_{\mathcal{S}_2}(\tilde{\bm v}) $, we have
% \begin{subequations}
%     \begin{align}
%         \|\tilde{\bm v}-\bm v \|_2^2 - \|\tilde{\bm v}-\bm v^\prime \|_2^2
%         =& 2\Re(\tilde{\bm v}^{\sf H}\bm v^\prime)-2\Re(\tilde{\bm v}^{\sf H}\bm v)\\
%         =&2\sum_{i=1}^{N}|\tilde{ v}_i|-2\Re\left(\sum_{i=1}^N |\tilde{v}_i|(v_i^\prime)^\ast v_i \right)\label{eq:3point1}\\
%         =& \sum_{i=1}^{N}|\tilde{ v}_i|\left(2-2\Re((v_i^\prime)^\ast v_i )\right)\\
%         =& \sum_{i=1}^{N}|\tilde{ v}_i|(v_i^\prime-v_i)^2\\
%         \ge& \min_i |\tilde{ v}_i| \| \bm v_i^\prime-\bm v_i \|_2^2,\label{eq:3point_s2}
%     \end{align}
% \end{subequations}
\begin{subequations}
    \begin{eqnarray}
    \lefteqn{\|\tilde{\bm v}-\bm v \|_2^2 - \|\tilde{\bm v}-\bm v^\prime \|_2^2}\nonumber \\ 
        &=& 2\Re(\tilde{\bm v}^{\sf H}\bm v^\prime)-2\Re(\tilde{\bm v}^{\sf H}\bm v)\\
        &=&2\sum_{i=1}^{N}|\tilde{ v}_i|-2\Re\left(\sum_{i=1}^N |\tilde{v}_i|(v_i^\prime)^\ast v_i \right)\label{eq:3point1}\\
        &=& \sum_{i=1}^{N}|\tilde{ v}_i|\left(2-2\Re((v_i^\prime)^\ast v_i )\right)\\
        &=& \sum_{i=1}^{N}|\tilde{ v}_i|(v_i^\prime-v_i)^2\\
        &\ge& \min_i |\tilde{ v}_i| \| \bm v_i^\prime-\bm v_i \|_2^2,\label{eq:3point_s2}
    \end{eqnarray}
\end{subequations}
where to write \eqref{eq:3point1} we use $\bm v^\prime = \Pi_{\mathcal{S}_2}(\tilde{\bm v})$, from which we can obtain $\tilde{v}_i = |\tilde{v}_i| v^\prime_i $. From \eqref{eq:3point_s2}, the three-point property of $\mathcal{S}_2$ is satisfied as long as $|\tilde{v}_i|$ is bounded away from 0, in which case the convergence guarantee can then be established. 
However, requiring that each element of $\Pi_{\mathcal{S}_1}(\bm v^{t}) $ is bounded away from zero for $t>t_0$ for some $t_0$ is a strong assumption. Nevertheless, it is uncommon for $\Pi_{\mathcal{S}_1}(\bm v^{t})$ to have near-zero elements; this may explain why the proposed alternating projection algorithm always converges from any random initial point in numerical simulations.

\subsection{Complexity Analysis}
At each iteration, the time complexity of the proposed alternating projection algorithm is dominated by the projection onto the set $\mathcal{S}_1$ in \eqref{eq:proj_s1}. Since we need to compute \eqref{eq:proj_s1} at each iteration, we can compute $\bm A^\ast (\bm A^\top \bm A^{\ast})^{-1}\bm A^\top$ at the first iteration and reuse its value for subsequent iterations. The complexity of computing $\bm A^\ast (\bm A^\top \bm A^{\ast})^{-1}\bm A^\top$ is $\mathcal{O}(N^2K^2+ K^6+K^4N)$ for general dense matrix $\bm A$. Thus, the total computational complexity is $\mathcal{O}(N^2K^2+ K^6+K^4N+tN^2)$, where $t$ is the number of iterations needed to converge. \re{If the alternating projection algorithm is initialized by the eigenvalue decomposition based method, additional $\mathcal{O}(N^2)$ time complexity is needed for computing the eigenvector associated with the largest eigenvalue \cite{trefethen1997numerical}. }

\section{Network Utility Maximization}\label{sect:utility}

The alternating projection algorithm seeks a zero-forcing solution to null 
the interference completely. However, except through the initialization method
as proposed in subsection \ref{subsection:initialization},  the iterates of
alternating projection do not attempt to maximize the desired signal power.
Thus, although the RIS is configured to suppress interference,
the zero-forcing solution does not necessarily maximize network utility. 
In this section, we consider the numerical optimization of network utility
functions in practical network settings. 

\subsection{Sum-Rate Maximization Algorithm}\label{sect:sum_rate}

Consider first the sum-rate maximization problem for the RIS system. We propose
the following two-stage algorithm that utilizes the zero-forcing solution to
maximize the sum rate in a $K$-user interference channel. First, we run the
alternating projection algorithm for the interference nulling problem. In the
second stage, the solution found by the alternating projection algorithm is
used as an initial point for the subsequent iterative algorithm for maximizing
the network sum rate. The second stage uses the RCG method in order to
accommodate the unit modulus constraints. The unit modulus constraints form a
Riemannian manifold \cite{yu2016mmWave, yu2019miso}.

More specifically, the sum-rate maximization problem is formulated as follows:
\begin{equation}\label{prob:sum_rate}
    \begin{aligned}
        &&\underset{\bm v}{\maxi}\quad &f_1(\bm v) \triangleq \sum_{k=1}^K R_k({\bm v})\\
        &&\subj\quad &|v_i| = 1, ~\forall i.
    \end{aligned}
\end{equation}
The above sum-rate maximization problem \eqref{prob:sum_rate} is nonconvex due
to the nonconvex objective function and nonconvex unit modulus constraints.
Since the objective function of problem \eqref{prob:sum_rate} is continuous and
differentiable, one straightforward approach is to use the projected gradient 
method. But instead of taking the Euclidean gradient, a better approach is to  
use the Riemannian gradient, which is formed by projecting the Euclidean
gradient onto the tangent space of the complex circle manifold formed by the
unit modulus constraints.  Further, to speed up convergence, the conjugate
Riemannian gradient may be used. This results in the RCG method, as proposed in 
\cite{yu2016mmWave, yu2019miso}.

The RCG method consists of three steps. First, the Euclidean gradient $\nabla f_1$ at the point $\bm v$ is computed as 
\begin{align}
    \nabla f_1(\bm v) = \sum_{k=1}^K \nabla R_k(\bm v),
\end{align}
where 
\begin{align}
    \nabla R_k(\bm v) = \frac{2 \sum_{j=1}^K p_j \bm a_{k,j}^\ast\bm a_{k,j}^\top\bm v}{\sum_{j=1}^K p_j|\bm a_{k,j}^{\top}\bm v|^2 +\sigma_0^2}- \frac{2 \sum_{j\neq k}p_j \bm a_{k,j}^\ast\bm a_{k,j}^\top\bm v}{\sum_{j\neq k} p_j|\bm a_{k,j}^{\top}\bm v|^2 +\sigma_0^2}.
\end{align}
By projecting the Euclidean gradient onto the tangent space of the complex unit circle, we obtain the Riemannian gradient as follows \cite{yu2019miso,guo2020weighted}:
\begin{align}\label{eq:R_gradient}
    \grad f_1(\bm v) = \mathcal{T}_{\bm v}(\nabla f_1(\bm v)),
\end{align}
where 
\begin{equation} \mathcal{T}_{\bm v}(\bm x)\triangleq\bm x -\Re(\bm x \circ \bm v^\ast) \circ \bm v
\end{equation} 
with $\circ$ denoting the element-wise product. 

Second, given the Riemannian gradient, the search direction $\bm d$ is updated by the conjugate gradient method as follows:
\begin{align}\label{eq:search_d}
    \bm d = -\grad f_1 + \lambda_1 \mathcal{T}_{\bm v}(\bar{\bm d}),
\end{align}
where $\lambda_1$ is the conjugate gradient update parameter, and $\bar{\bm d} $ is the previous search direction.

Finally, to keep the updated point on the manifold, we need a retraction operation to project the point onto the manifold as follows:
\begin{align}\label{eq:rcg_update}
    \bm v \leftarrow (\bm v+\lambda_2 \bm d)/|\bm v+\lambda_2 \bm d|,  
\end{align}
where $\lambda_2$ is the Armijo step size. 

\begin{algorithm}[t]
    \caption{Two-Stage Algorithm for Solving the Sum-Rate Maximization Problem \eqref{prob:sum_rate}}\label{algo:sum-rate}
    \KwIn{Vector $\bm v\in\mathbb{C}^N$ for initializing Algorithm~\ref{algo:alter}. Channel matrix $\bm A$.}
    {\bf First Stage}: Run Algorithm~\ref{algo:alter} from $\bm v$\\
    % \vspace{0.3cm}
    {\bf Second Stage}: Set $\bm v^0$ to be the output of Algorithm~\ref{algo:alter} \\
    \For{$t=0, 1,2,\cdots$}{
        Compute $\grad f_1(\bm v^t)$ by \eqref{eq:R_gradient}\\
        Compute search direction $\bm d$ by \eqref{eq:search_d}\\
        Update $\bm v^{t+1}$ by \eqref{eq:rcg_update}\\
        \If{stopping criterion is satisfied}{break}
    }
    \KwOut{$\bm v^{t+1}$}
\end{algorithm}

The overall algorithm is summarized as Algorithm \ref{algo:sum-rate}.
The RCG method is guaranteed to converge to a stationary point of the problem \eqref{prob:sum_rate} \cite{absil2009optimization}. 
The main advantage of using the zero-forcing solution as the initial point is that it helps the RCG method avoid undesirable local maxima. The numerical results in Section~\ref{sect:simulations} show that such a two-stage optimization algorithm achieves significantly better performance as compared to methods with random initializations.  

\subsection{Minimum-Rate Maximization Algorithm}\label{sect:minimum_rate}
The sum-rate objective function does not take fairness into consideration. In this section, we investigate how to design the RIS system in order to maximize the minimum rate over all links in a $K$-user interference channel. 

The minimum-rate maximization problem can be formulated as the following optimization problem:
\begin{equation}\label{prob:mim_rate}
    \begin{aligned}
        &&\underset{\bm v}{\mini}\quad &f_2(\bm v)\triangleq \max_k \{-R_k\}\\
        &&\subj\quad &|v_i| = 1,~\forall i.\\
    \end{aligned}
\end{equation}
Compared with the sum-rate maximization problem \eqref{prob:sum_rate}, 
the objective function of the minimum rate maximization problem
\eqref{prob:mim_rate} is no longer smooth, so RCG or gradient-based methods
cannot be directly applied.  Existing methods in the literature include
bisection or fractional programming with SDR 
\cite{alwazani2020intelligent,9246254}, or penalty method followed by SCA
\cite{yu2020joint}.

This paper proposes a subgradient projection method to solve the problem
\eqref{prob:mim_rate}. This method is found to have good performance and has a
complexity which is scalable even for large values of $N$.  We numerically show
in Section~\ref{sect:simulations} that the subgradient method can outperform
the SDR-based method considerably, and has complexity much lower than SCA.

{A vector  $\bm g(\bm v^\prime) \in\mathbb{C}^N$ is a subgradient of the function $f_2(\bm v^\prime)$ if it is an element of the Fr\'echet subdifferential  $\partial f_2(\bm v^\prime) \triangleq\left\{g(\bm v^\prime)|~\underset{\bm v\rightarrow \bm v^\prime}{\liminf}  \frac{f_2(\bm v)-f_2(\bm v^\prime)-2\Re(\langle g(\bm v^\prime), \bm v-\bm v^\prime\rangle )}{\|\bm v-\bm v^\prime\|_2 }\ge 0\right\}$ \cite{rockafellar2009variational}. Next, we show that a subgradient of $f_2(\bm v^\prime)$ can be given as
\begin{equation}\label{eq:subgradient}
    \bm g (\bm v^\prime)= -\nabla R_i(\bm v^\prime),
\end{equation} 
where $i = {\rm argmax}_k \{-R_k(\bm v^\prime)\}$. The proof follows the proof technique from \cite{zhang2021fast}.  Since $\mathcal{S}_2$ is compact, $-\nabla R_k(\bm v)$ is finite for any $\bm v\in\mathcal{S}_2$ and $k=1,\cdots,K$. Then there exists a constant $L>0$ such that $\|\nabla R_k(\bm v)-\nabla R_k(\bm v^\prime)\|_2\le L\|\bm v-\bm v^\prime\|_2$. This implies $-R_k(\bm v)$ is $L$-smooth. Thus for all $k$ and for all $\bm v, \bm v^\prime\in\mathcal{S}_2$, we have \cite{beck2017first}
\begin{equation}
    -R_k(\bm v)\ge -R_k(\bm v^\prime)+ 2\Re(\langle -\nabla R_k(\bm v^\prime),\bm v-\bm v^\prime\rangle)-\frac{L}{2}\|\bm v-\bm v^\prime\|_2^2. \notag
\end{equation}
 Let $i = {\rm argmax}_k \{-R_k(\bm v^\prime)\}$, we thus have 
\begin{align}
    f_2(\bm v)&\ge -R_i(\bm v)\notag\\
    &\ge -R_i(\bm v^\prime)+ 2\Re(\langle -\nabla R_i(\bm v^\prime),\bm v-\bm v^\prime\rangle)-\frac{L}{2}\|\bm v-\bm v^\prime\|_2^2\notag\\
    &=f_2(\bm v^\prime)+2\Re(\langle -\nabla R_i(\bm v^\prime),\bm v-\bm v^\prime\rangle)-\frac{L}{2}\|\bm v-\bm v^\prime\|_2^2.\label{eq:proof_subgradient}
\end{align}
By rearranging the terms in \eqref{eq:proof_subgradient} and taking $\liminf$ for $\bm v\rightarrow \bm v^\prime$, we conclude that \eqref{eq:subgradient} is a subgradient of $f_2(\bm v^\prime)$.  
}

Given the subgradient \eqref{eq:subgradient}, the projected subgradient method takes the following form:
\begin{align}\label{eq:iter_min}
    \bm v^{t+1} = \Pi_{\mathcal{S}_2} \left(\bm v^{t}-\gamma^{t+1}\bm g (\bm v^t)\right),
\end{align}
where $\gamma^{t+1}$ is the step size at the $(t+1)$-th iteration.  The step size of the projected subgradient method needs to be properly chosen to promote fast convergence. One example of step size selection is to choose the constant step size normalized by the norm of the subgradient \cite{boyd2003subgradient}, i.e., 
\begin{align}
    \gamma^{t+1} = \frac{c}{\|\bm g(\bm v^t)\|_2},
\end{align}
which is shown to achieve good performance in practice. Because the subgradient method cannot guarantee decrease in the objective functions for each iteration, we keep track of the best point found so far during the iterations.

Unlike the sum-rate maximization problem, the initialization by the
zero-forcing solution does not have a significant impact for the performance 
of the minimum-rate maximization problem. This is because the minimum rate objective function
typically has a bottleneck link, and it is only crucial to null the interference
for the bottleneck link, but not the other links.

\section{RIS System with Direct Paths Between the Transmitters and Receivers}\label{sec:Direct}

So far, we have considered the system model where the direct paths between the transmitters and the receivers are blocked. However, the algorithms proposed in this paper are not limited to this scenario.
In this section, we extend the proposed algorithms to the RIS systems with direct paths.

Let $b_{k,j}$ denote the direct channel between transmitter $j$ and receiver $k$, the received signal model \eqref{eq:channel_model} can be rewritten as
\begin{align}\label{eq:received_y_k_direct}
    y_k = (\bm a_{k,k}^{\top}\bm v +b_{k,k}) s_k+\sum_{j\neq k}  (\bm a_{k,j}^{\top} \bm v + b_{k,j}) s_j+n_k.
\end{align}
In this case, the achievable rate of the $k$-th link becomes
\begin{align}\label{eq:rate_direct}
    R_k= \log_2\left(1+\frac{| \bm a_{k,k}^{\top}\bm v +b_{k,k}|^2 p_k}{\sum_{j\neq k} |\bm a_{k,j}^{\top} \bm v + b_{k,j}|^2 p_j+\sigma_0^2}\right).
\end{align}

Similar to the interference nulling conditions in \eqref{eq:zf_cond00}, \eqref{eq:zf_cond11}, \eqref{eq:zf_cond12}, we can establish the interference nulling conditions for the case with direct paths as follows:
\begin{subequations}
    \begin{align}
	    \bm a_{k,k}^{\top}\bm v +b_{k,k} \neq 0, \quad &k=1,\cdots,K, \label{eq:zf_cond00_direct}\\
	    \bm a_{k,j}^{\top}\bm v +b_{k,j} = 0, \quad &k=1, \cdots, K, \ \ \forall j\neq k, \label{eq:zf_cond11_direct}\\
	    |v_i| = 1, \quad &i=1,\cdots,N.\label{eq:zf_cond12_direct}
    \end{align}
\end{subequations}
Accordingly, the problem \eqref{prob:IA_feasibility_1} can be reformulated as that of finding a feasible solution to the following problem:
\begin{equation}\label{prob:IA_feasibility_1_direct}
    \begin{aligned}
        &~~~\operatorname{find}\quad && \bm v \\
        &\subj\quad &&\bm  A^\top \bm v +\bm b = \bm  0 \\
        &\quad&&|v_i|= 1,~i=1,\cdots,N.
    \end{aligned}
\end{equation} 

The proposed alternating projection algorithm can be readily applied to solve the above problem \eqref{prob:IA_feasibility_1_direct}. Specifically, the set $\mathcal{S}_1$ in \eqref{eq:omega_1} becomes an affine set $\mathcal{S}_1 =\{ \bm v: \bm  A^\top \bm v +\bm b = \bm  0\} $, and the projection to this affine set $\mathcal{S}_1$ also has
a closed-form solution \cite{parikh2014proximal}:
\begin{align}
    \Pi_{\mathcal{S}_1}(\bm v) &= \bm v - \bm A^\ast (\bm A^\top \bm A^{\ast})^{-1}(\bm A^\top \bm v+\bm b).
\end{align}
The other parts of the proposed alternating projection algorithm remain unchanged. The local convergence guarantee in Theorem \ref{theorem:ap_convergence} still holds since the new affine set $\mathcal{S}_1$ remains convex.

We note that the strength of the direct channel $\bm b$ affects the feasibility of the problem \eqref{prob:IA_feasibility_1_direct}. For a zero-forcing solution to exist, it is necessary to have 
\begin{align}\label{eq:neccesary_direct_link}
    \|\bm a_{k,j} \|_1 \ge |b_{k,j}|, \quad \forall k=1,\cdots,K, \ \ j\neq k,
\end{align}
otherwise even aligning the strength of all the cascaded channels, i.e., all the elements $\bm a_{k,j}$, cannot cancel out the interference caused by the direct channel $b_{k,j}$. \re{From \eqref{eq:neccesary_direct_link}, the value of $ \|\bm a_{k,j} \|_1 = \sum_{i=1}^N |[\bm a_{k,j}]_i|$ depends on both the strength of the cascaded channel $|[\bm a_{k,j}]_i|$ and the number of RIS elements $N$. 
If the strength of the cascaded channel $|[\bm a_{k,j}]_i|$ is relatively small, 
e.g., due to the large loss caused by the passive reflection \cite{9306896},
as compared to the direct channel $|b_{k,j}|$, 
the number of RIS elements $N$ would need to be large in order to 
cancel the interference caused by both the direct and the reflection paths.}

The algorithms developed for maximizing the sum rate or the minimum rate can be likewise extended to the case with direct paths by recomputing the gradients as
\begin{align}
    \nabla R_k(\bm v) = &\frac{2 \sum_{j=1}^K p_j\bm a_{k,j}^\ast\bm a_{k,j}^\top\bm v+2\sum_{j=1}^K p_j\bm a_{k,j}^\ast b_{k,j}}{\sum_{j=1}^K p_j|\bm a_{k,j}^{\top}\bm v +b_{k,j}|^2 +\sigma_0^2}\notag\\
    &-\frac{2 \sum_{j\neq k}p_j\bm a_{k,j}^\ast\bm a_{k,j}^\top\bm v+2\sum_{j\neq k} p_j\bm a_{k,j}^\ast b_{k,j}}{\sum_{j\ne k} p_j|\bm a_{k,j}^{\top}\bm v +b_{k,j}|^2 +\sigma_0^2}.
\end{align}

\section{Simulation Results}\label{sect:simulations}

In this section, we present numerical results to evaluate the performance of the proposed algorithms for optimizing the RIS system with different objective functions. We first evaluate the algorithm for the scenario without direct paths, then consider the scenario with direct links. 

\subsection{Simulation Setup}
\begin{figure}[t]
    \centering
    \includegraphics[width=9cm]{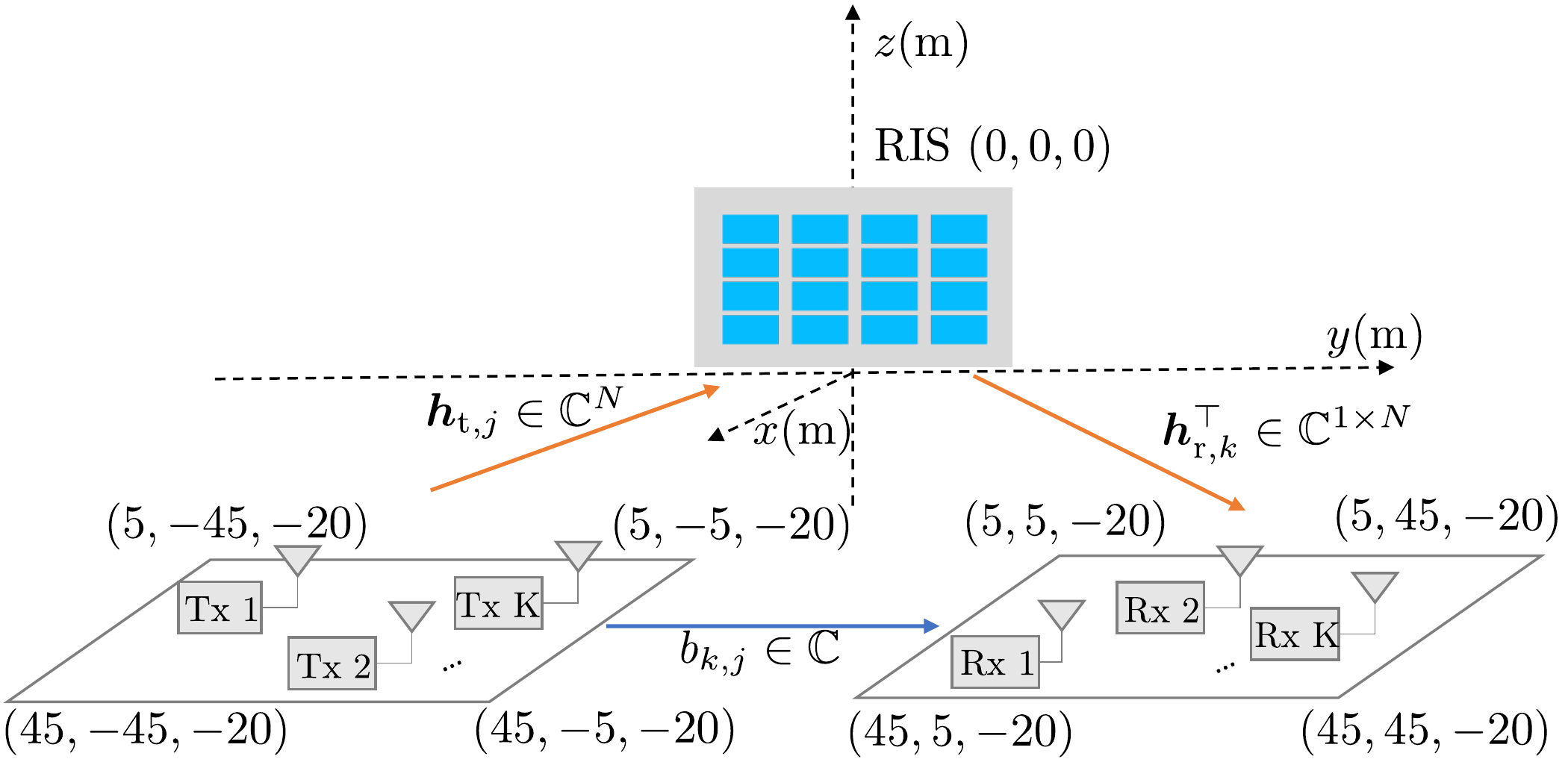}
    \caption{The simulation scenario.}
    \label{fig:simulation_setup}
\end{figure}

We consider a simulation setup as shown in Fig.~\ref{fig:simulation_setup}. An RIS equipped with a uniform rectangular array is placed on the $(y,z)$-plane at the coordinate $(0,0,0)$ in meters. The transmitters and the receivers are uniformly distributed in a rectangular area $[5,45]\times[-45,-5]$ and $[5,45]\times[5,45]$ respectively, and the $z$-coordinates of all the users are set to be $z=-20$.  The channel model between the RIS and the user $k$ is assumed to be Rician fading channel as in \cite{9427148}, which is modeled as
\begin{align}\label{eq:rician_channel}
    \bm h_k = \sqrt{\varepsilon \over {1+\varepsilon}} \beta_k\tilde{\bm h}_k^{\rm LOS}+ \sqrt{1 \over {1+\varepsilon}}\beta_k\tilde{\bm h}_k^{\rm NLOS},
\end{align}
where $\beta_k$ is the path-loss from the RIS to the user $k$, $\tilde{\bm h}_k^{\rm LOS}$ is the line-of-sight part and $\tilde{\bm h}_k^{\rm NLOS}$ is the non-line-of-sight part. The path-loss in dB is modeled as $-30-22\log(d_k)$ where $d_k$ is the distance between the user $k$ and the RIS in meters. The line-of-sight channel $\tilde{\bm h}_k^{\rm LOS}$ is modeled as in \eqref{eq:channel_model_ray}, i.e., $\tilde{\bm h}_k^{\rm LOS} = \tilde{\bm a}(\theta_k,\phi_k)$ and the entries of the non-line-of-sight channel vector $\tilde{\bm h}_k^{\rm NLOS}$ are modeled as i.i.d. standard Gaussian distributions, i.e., $\left[\tilde{\bm h}_k^{\rm NLOS}\right]_i\sim\mathcal{CN}(0,1)$. The system bandwidth is $10$MHz and the noise spectral density is $-170$dBm/Hz. 

\re{In Section \ref{subsec:int_null} to \ref{subsec:mim_rate_sim}, we consider the scenario without the direct paths, i.e., $b_{k,j}=0$. The scenario with the direct paths is considered in Section \ref{subsec:sim_direct}, where the model for the direct channels is described in detail.}
\re{\begin{table}[t]
    \caption{Simulation Parameters}
    \centering
    \begin{tabular}{ |c|c| } 
     \hline
     RIS location & $(0\text{m},0\text{m},0\text{m})$  \\
     \hline
     Tx location & $[5\text{m},45\text{m}] \times[-45\text{m},-5\text{m}]$ \\ 
     \hline
     Rx location & $[5\text{m},45\text{m}] \times[5\text{m},45\text{m}]$ \\ 
     \hline 
     Path-loss for $\bm h_{{\rm t},j}$ and $\bm h_{{\rm r},k}$ & $-30-22\log(d)$ \\ 
     \hline 
     Bandwidth & $10$MHz  \\ 
     \hline
     Noise power spectral density & $-170$dBm/Hz\\
     \hline
    \end{tabular}
\end{table}}

\subsection{Interference Nulling Problem}\label{subsec:int_null}

For the interference nulling problem,  to verify the effectiveness of the proposed alternating projection algorithm, we compare it to the following projected gradient method as a baseline.

\subsubsection{Baseline Projected Gradient Method}
The interference nulling problem \eqref{prob:IA_feasibility_1} can also be formulated as follows:
\begin{equation}\label{prob:IA_feasibility_proj}
    \begin{aligned}
        &\underset{{\bm v}}{\mini}\quad &&f_3(\bm v)\triangleq\|\bm  A^\top \bm v\|_2^2 \\
        &\subj\quad && |v_i|= 1,~i=1,\cdots,N,\\
    \end{aligned}
\end{equation} 
where the sum of interference is minimized subject to the unit modulus constraint on $\bm v$. If the objective function of problem \eqref{prob:IA_feasibility_proj} is minimized to be zero, then we find a zero-forcing solution that nulls all the interference. The problem \eqref{prob:IA_feasibility_proj} can be solved by the projected gradient method \cite{7849224}, which has the following update rule:
\begin{align}
    &\tilde{\bm v}^{t} = \bm v^{t} - \gamma^t\cdot 2 \bm A^\ast\bm A^\top\bm v^t, \label{eq:pgd_ia1}\\
    &\bm v^{t+1} = \Pi_{\mathcal{S}_2}(\tilde{\bm v}^{t}), \label{eq:pgd_ia2}
\end{align}
where $\gamma^t$ is the step size at the $t$-th iteration. The choice of an appropriate step size is crucial for the algorithm to converge. However, it is computationally expensive to identify the optimal step size that minimizes the objective function along the descent direction. A common practice is to select the step size by backtracking line search \cite{boyd2004convex}. Specifically, let $\gamma^t=1$,  $\alpha\in(0,0.5)$ and $\beta\in(0,1)$. We keep shrinking the step size by $\gamma^t=\beta \gamma^t$ until $f_3(\bm v^t-\gamma^t\nabla f_3(\bm v^{t}))\le f_3(\bm v^t)-\alpha\lambda\|\nabla f_3(\bm v^t) \|_2^2$.

We should note that the proposed alternating projection method is a parameter-free algorithm, while the projected gradient method needs to adjust its step size at each iteration.

\begin{figure}[t]
    \centering
    \includegraphics[width=8.6cm]{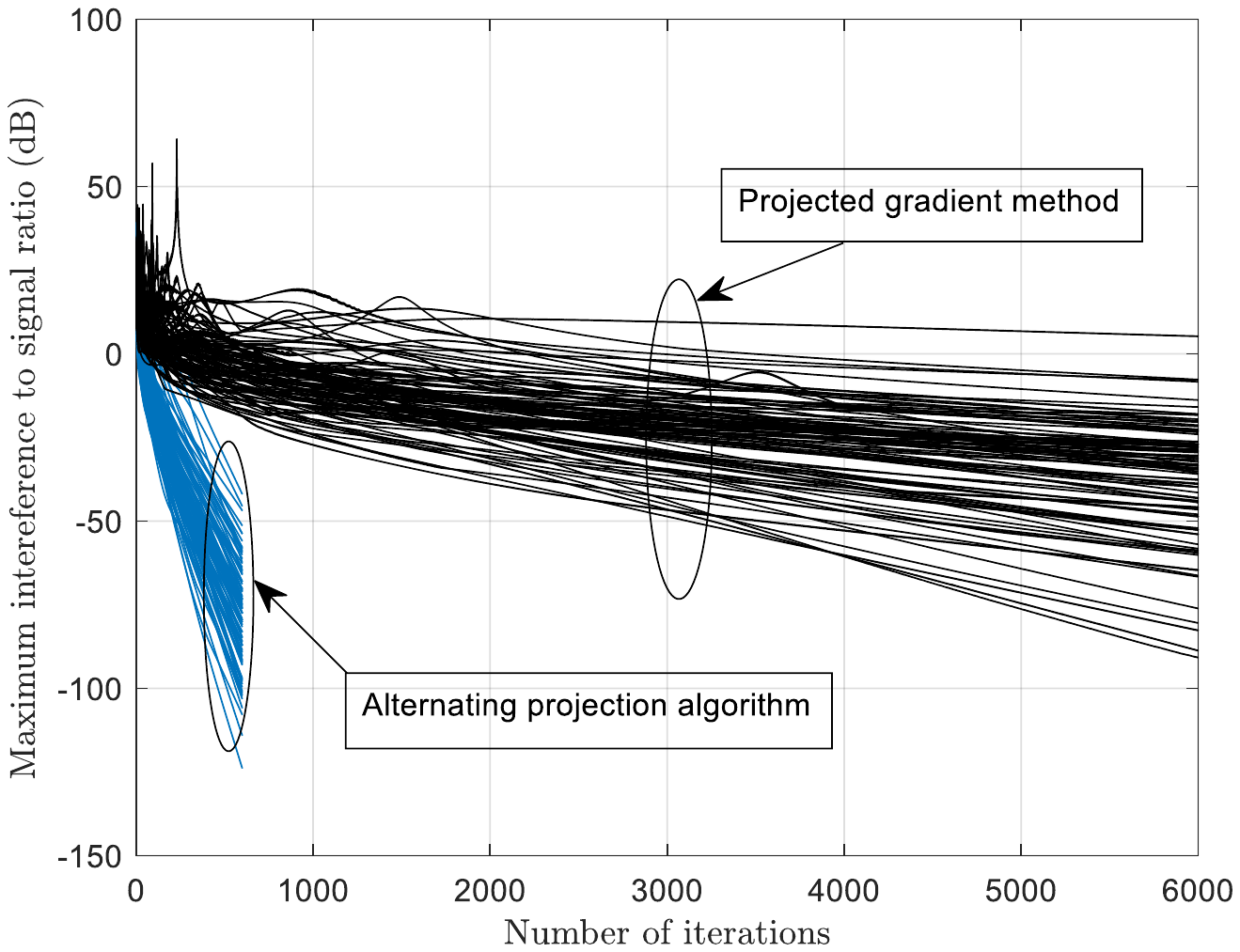}
    \caption{Convergence of the proposed alternating projection Algorithm~\ref{algo:alter} \changeTao{vs.} the projected gradient method. The number of RIS elements is $N = 12\times 12$ and the number of transceiver pairs is $K=8$.}
    \label{fig:main_AP_convergence}
\end{figure}

\begin{figure*}[t]
    \centering
    \subfigure[Empirical interference nulling probability vs. number of RIS elements. \label{fig:phase_transition2}]{\includegraphics[width=8.56cm]{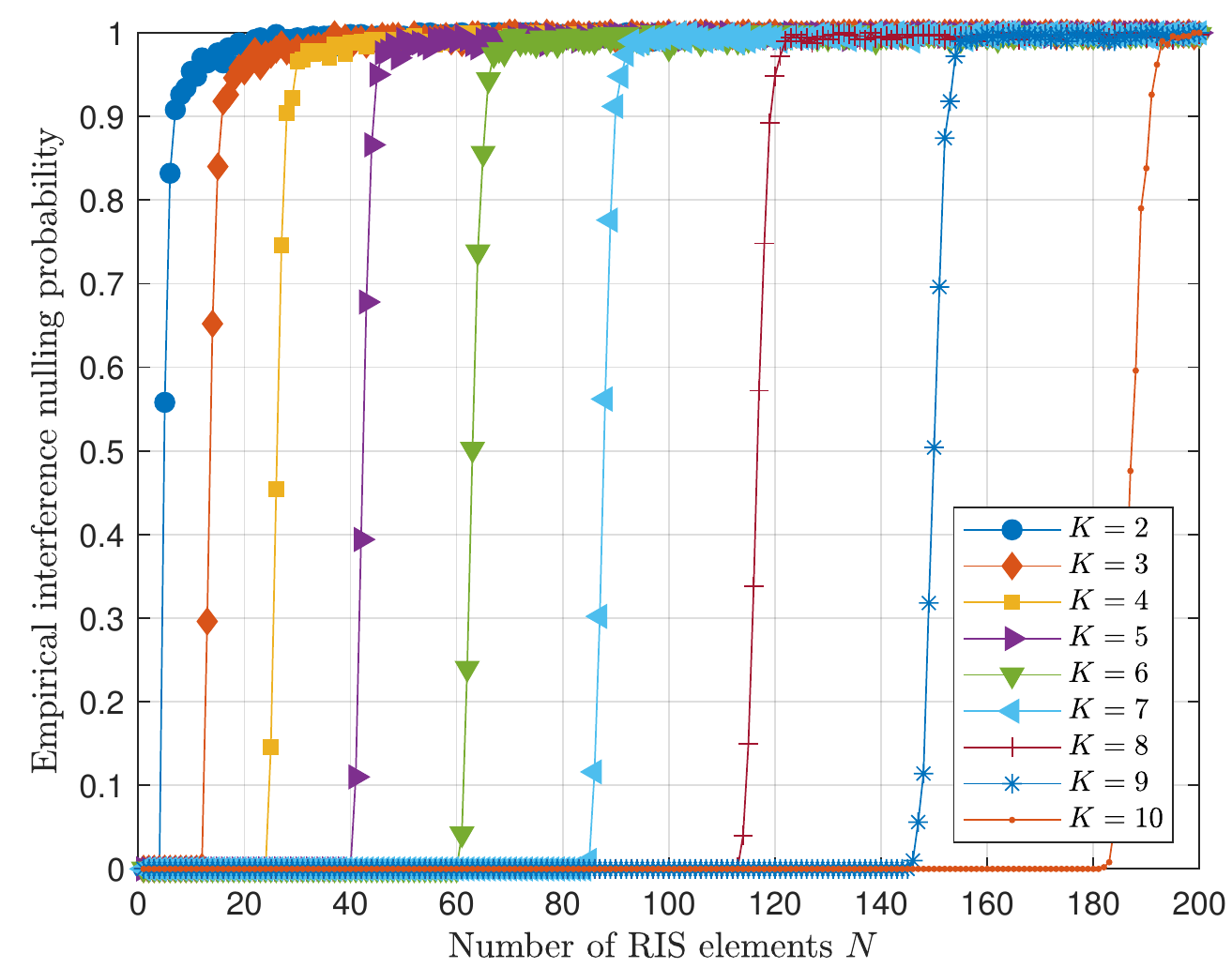}}
    \subfigure[Number of RIS elements $N$ vs. number of transceiver pairs $K$.\label{fig:phase_transition1}]{\includegraphics[width=8.6cm]{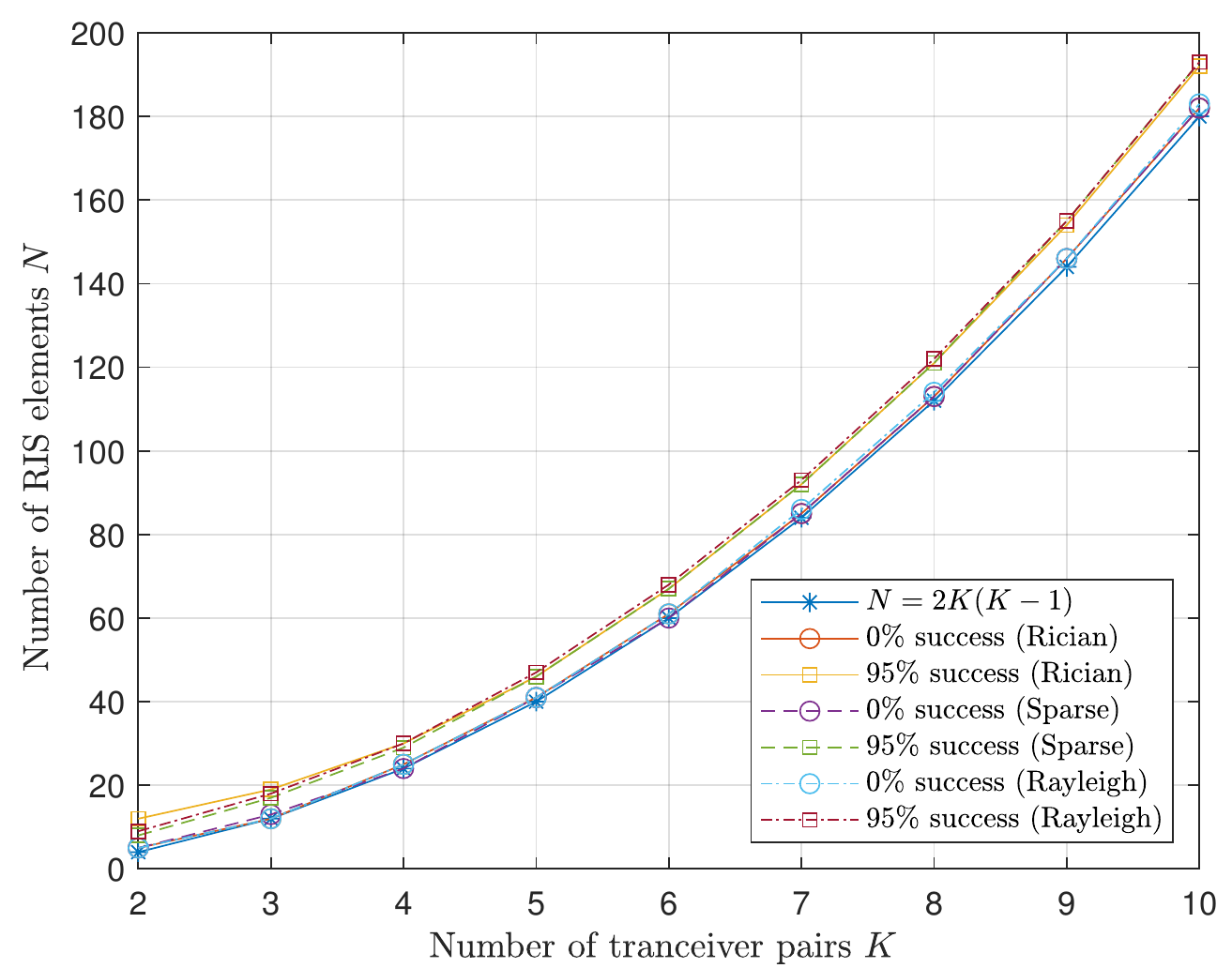}}
    \caption{Phase transition phenomenon for interference nulling in RIS-enabled $K$-user interference channels without direct links.}    
\end{figure*}

\subsubsection{Convergence Speed}

We now provide simulations to evaluate the convergence rate of the proposed
alternating projection method, i.e., Algorithm~\ref{algo:alter}, as compared to
the projected gradient method. In this experiment, the RIS consists of
$12\times 12$ passive elements and serves $8$ transceiver pairs. We run $100$
independent random trials with different random initial points for the proposed
alternating projection algorithm and the projected gradient method. The maximum
interference-to-signal ratio (ISR) over the $K$ links is adopted as 
the performance metric. % as shown in Fig.~\ref{fig:main_AP_convergence}. 
For the projected gradient method, the backtracking search parameters are set as
$\alpha=0.3$ and $\beta=0.8$. 

From Fig.~\ref{fig:main_AP_convergence}, it can be observed that the alternating projection algorithm can decrease the maximum ISR  to about $-50$dB within $600$ iterations for almost all the trials. However, the projected gradient method takes around $6000$ iterations to reach this level of maximum ISR. We observe that the maximum ISR is not always monotonically decreasing at the beginning. This is because the alternating projection algorithm and the projected gradient method only attempt to reduce the interference, without regard to the power of the useful signals, so it does not necessarily result in a decrease in the maximum ISR. However, the maximum ISR eventually decreases almost linearly in later iterations. %, implying that reducing interference eventually has a dominant impact on the ISR. 

\subsubsection{Phase Transitions}

To evaluate the performance of the proposed alternating projection algorithm
for finding a zero-forcing solution, we also adopt the maximum ISR as the
metric. We declare that an interference nulling solution is found if the
maximum ISR across all transceiver pairs is below $-60$dB.  We evaluate
systems with $K$ ranging from $2$ to $10$ and $N$ ranging from $1$ to $200$. 
The RIS is assumed to be a uniform linear array for convenience so that we can
increase $N$ at a step size $1$.  For each setting of $K$ and $N$, $500$
independent trials are conducted to estimate the
interference nulling probability. 

To evaluate the impact of different channel
distributions on the performance, we also consider the sparse channel model and
Rayleigh fading channel model. For the sparse channel model, 
$ \bm h_k = \frac{1}{\sqrt{L}}\sum_{\ell=1}^L\alpha^\ell\tilde{\bm a}(\theta^\ell_k,\phi^\ell_k)$, where the superscript $\ell$ denotes the
$\ell$-th path, with the number of path $L=5$ for the channels between the transmitters/receivers and the RIS. For the Rayleigh fading channel
model, we have that $\bm h_k \in\mathcal{CN}(\bm 0, \bm I)$ for the channels
between the transmitters/receivers and the RIS.

In Fig.~\ref{fig:phase_transition2}, we plot the empirical success probability
versus the number of RIS elements $N$ for the Rician channel model as in
\eqref{eq:rician_channel}. We observe a phase transition phenomenon, i.e., 
for any fixed $K$, the empirical interference nulling probability transitions 
sharply from $0$ to $1$ as $N$ exceeds a threshold. As $K$ increases, 
the phase transition location grows approximately as $2K(K-1)$, implying that
the interference can be completely nulled with high probability using the
proposed alternating projection algorithm if $N$ is slightly greater than
$2K(K-1)$. 

The phase transition location can be more precisely observed in
Fig.~\ref{fig:phase_transition1}.  In Fig.~\ref{fig:phase_transition1}, the
points marked as ``$0\%$ success'' refer to the points below which all the
trials fail to find a zero-forcing solution. The points marked as ``$95\%$
success'' refer to the points above which 
a zero-forcing solution can be found with empirical probability greater than $95\%$.
From Fig.~\ref{fig:phase_transition1}, we can see the that ``$0\%$ success''
lines coincide with the line $N=2K(K-1)$, which implies that the
$N\ge2K(K-1)$ is a necessary condition for the existence of a zero-forcing
solution. It is also observed that if $N$ is only slightly larger than
$2K(K-1)$, the proposed alternating projection algorithm can already find 
a zero-forcing solution with high probability. Furthermore, the distribution of
the channels does not affect the performance of the proposed alternating
projection algorithm or the location of phase transition.

\subsection{Sum-Rate Maximization}

\begin{figure*}[t]
    \centering
    \subfigure[Sum rate vs. transmit power. The number of RIS elements is $N=12\times 12$ and the number of \changeTao{transceiver} pairs is $K=8$. \label{fig:sumrate_a}]{\includegraphics[width=8.5cm]{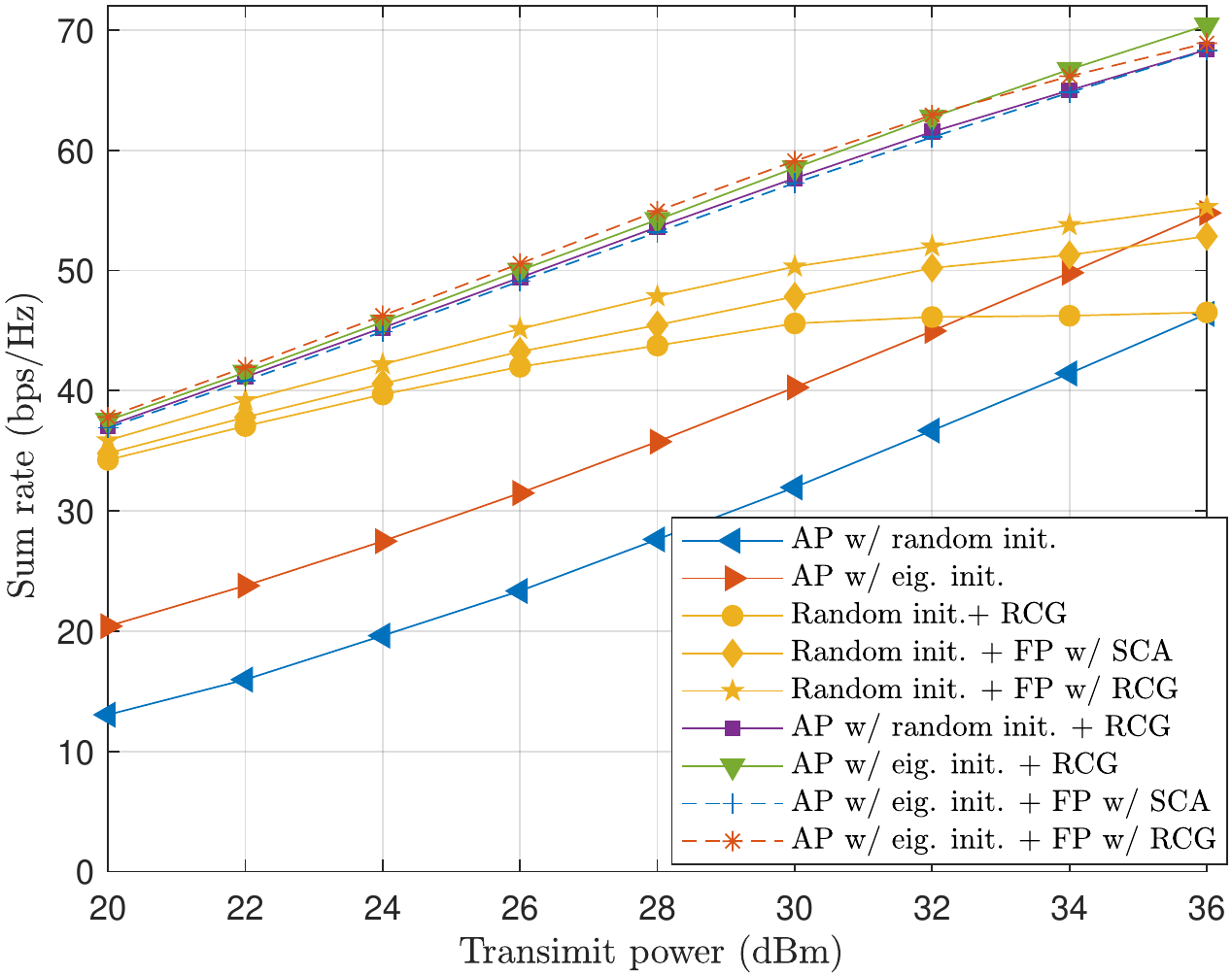}}
    \hspace{2pt}
    \subfigure[Sum rate vs. number of RIS elements. The number of transceiver pairs is $K=6$  and the transmit power is $35$dBm. \label{fig:sumrate_b}]{\includegraphics[width=8.6cm]{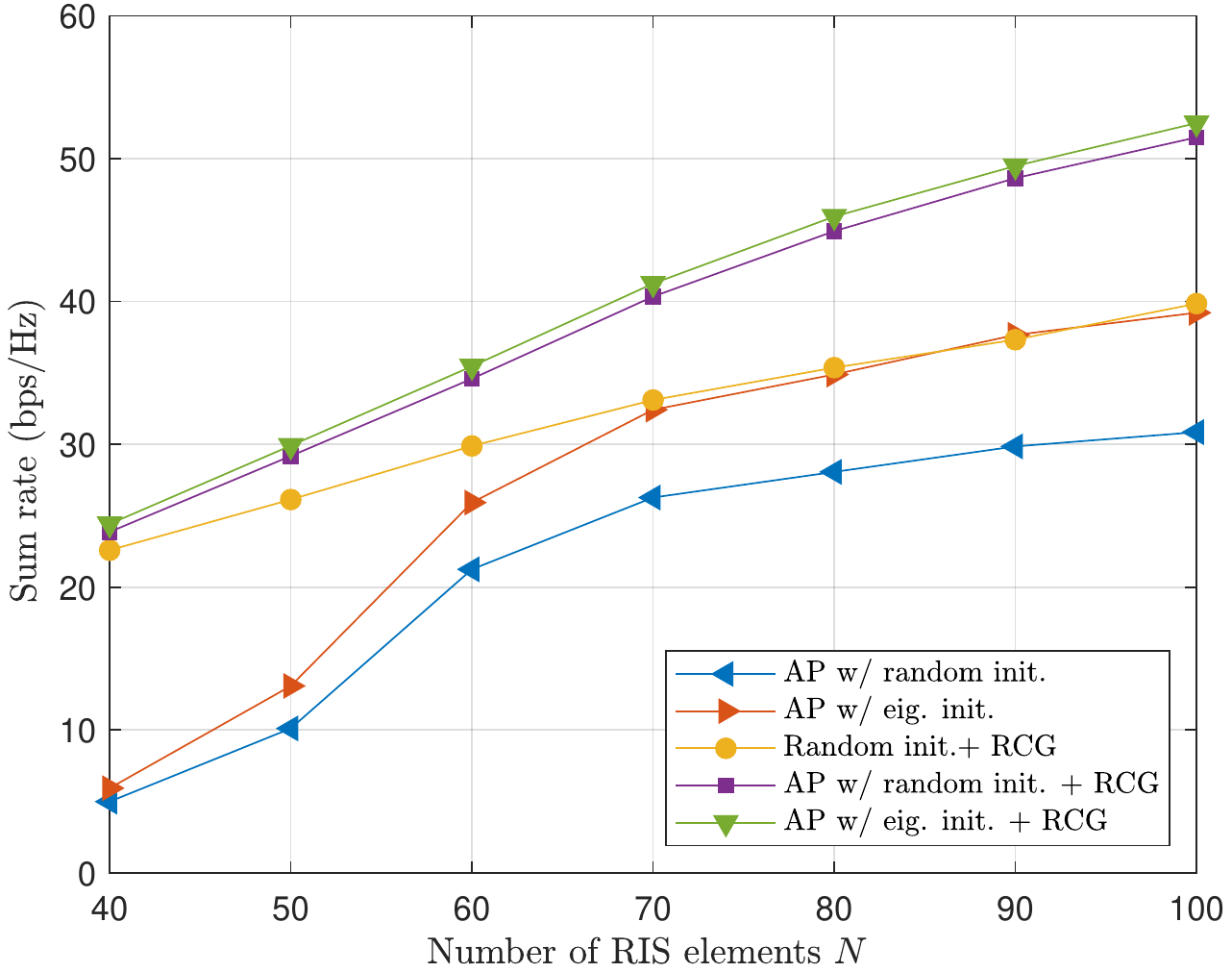}}
    \caption{Sum rate maximization performance of an RIS-enabled interference channel without direct links. }
    \label{fig:sumrate}
\end{figure*}

To demonstrate the efficiency of the proposed two-stage method for solving the sum-rate maximization problem \eqref{prob:sum_rate}, we compare the sum rate achieved by the following schemes:
\begin{itemize}
    \item \textit{AP With Random Init.}:  The alternating projection algorithm for zero-forcing is initialized by a unit modulus vector $\bm v\in\mathbb{C}^N$ with random phases.
    \item \textit{AP With Eigen Init.}:  The alternating projection algorithm for zero-forcing is initialized by the eigenvalue decomposition scheme proposed in Section~\ref{subsection:initialization}.
    \item \textit{Random Init. + RCG}: The RCG algorithm for sum-rate maximization is initialized by a unit modulus vector $\bm v\in\mathbb{C}^N$ with random phases. 
    \item \textit{AP With Random Init. + RCG}:  The RCG algorithm for sum-rate maximization is initialized by a zero-forcing solution obtained by the alternating projection algorithm, which is itself initialized by a unit modulus vector $\bm v\in\mathbb{C}^N$ with random phases.
    \item \textit{AP With Eigen Init. + RCG}: The RCG algorithm for sum-rate maximization is initialized by a zero-forcing solution obtained by the alternating projection algorithm, which is itself initialized by the eigenvalue decomposition scheme proposed in Section~\ref{subsection:initialization}.
\end{itemize}  

\re{
Moreover, we compare the performance of directly using RCG on the objective
function to an alternative approach of first reformulating the sum-rate
maximization problem using a closed-form fractional programming (FP) technique \cite{8310563}, then
solving the reformulated problem by block coordinate descend (BCD).  
In each step of the BCD algorithm, the reflection coefficients at the RIS 
can be updated by either SCA \cite{guo2020weighted} or RCG. 
The zero-forcing solution obtained by alternating projection can also be 
used as the initialization point in this case. Specifically, the following
schemes are compared: }
\re{\begin{itemize}
    \item \textit{Random Init. + FP With SCA}: The FP based algorithm with SCA is initialized by a unit modulus vector $\bm v\in\mathbb{C}^N$ with random phases. 
    \item \textit{Random Init. + FP With RCG}: The FP based algorithm with RCG is initialized by a unit modulus vector $\bm v\in\mathbb{C}^N$ with random phases. 
    \item \textit{AP With Eigen Init. + FP With SCA}: The FP based algorithm with SCA is initialized by the zero-forcing solution obtained by the proposed alternating projection algorithm with the eigenvalue decomposition initialization scheme.
    \item \textit{AP With Eigen Init. + FP With RCG}: The FP based algorithm with RCG is initialized by the zero-forcing solution obtained by the proposed alternating projection algorithm with the eigenvalue decomposition initialization scheme. 
\end{itemize}
}

The simulation results of the above different approaches are shown in
Fig.~\ref{fig:sumrate}, where each point is averaged over $500$ independent
channel realizations.
For the experiments in Fig.~\ref{fig:sumrate_a}, we set the number of RIS
elements and the number of transceiver pairs respectively to $N=12\times 12$
and $K=8$. \re{As can be seen from Fig.~\ref{fig:sumrate_a}, with the random
initialization, the sum rates achieved by the RCG method and the FP based
method increase with the transmit power initially with FP outperforming RCG, 
but their performances gradually saturate as the transmit power increases.  On
the other hand, although the sum rates achieved by the zero-forcing solutions
are lower at low SNR, their performances increase almost linearly as
the transmit power increases, and eventually catch up with the RCG or FP
algorithms at high SNR. Between the two zero-forcing schemes, 
using the eigenvalue decomposition based initialization
significantly outperforms random initialization. But the best performance is
achieved if the RCG or the FP based method is initialized by the zero-forcing
solution. Such a two-stage optimization scheme can achieve much better
performance than using either the zero-forcing solution directly or the RCG 
or FP based method with random initialization.
}

\re{
We note that although the FP based algorithm outperforms RCG if random
initialization is used, their performances are almost indistinguishable when
the zero-forcing solution is used as the initialization. In terms of complexity,
using the RCG algorithm directly on the sum-rate objective has a lower
complexity than FP, because an iterative algorithm (SCA or RCG) is needed to
update the reflection coefficients at each iteration of the BCD algorithm after
the FP step. Thus, using the RCG algorithm with zero-forcing initialization
is overall the best approach. 
}

In Fig.~\ref{fig:sumrate_b}, we plot the sum rate versus the number of RIS elements for fixed  $35$dBm transmit power with the number of transceivers $K=6$. The RIS consists of $10$ elements in the horizontal direction.  From Fig.~\ref{fig:sumrate_b}, we can observe that RCG with the zero-forcing solution as the initial point consistently achieves the best performance as compared to the other schemes. We observe a significant increase in the sum rate for the alternating projection algorithm when $N$ changes from $50$ to $60$. This is because $60$ is around the phase transition point when $K=6$ (see Fig.~\ref{fig:phase_transition2}). It is interesting to observe that RCG with the solution returned by the alternating projection algorithm as the initial point can outperform RCG with a random initialization scheme even if the interference cannot be suppressed completely.

\begin{figure}[t]
    \centering
	\includegraphics[width=8.6cm]{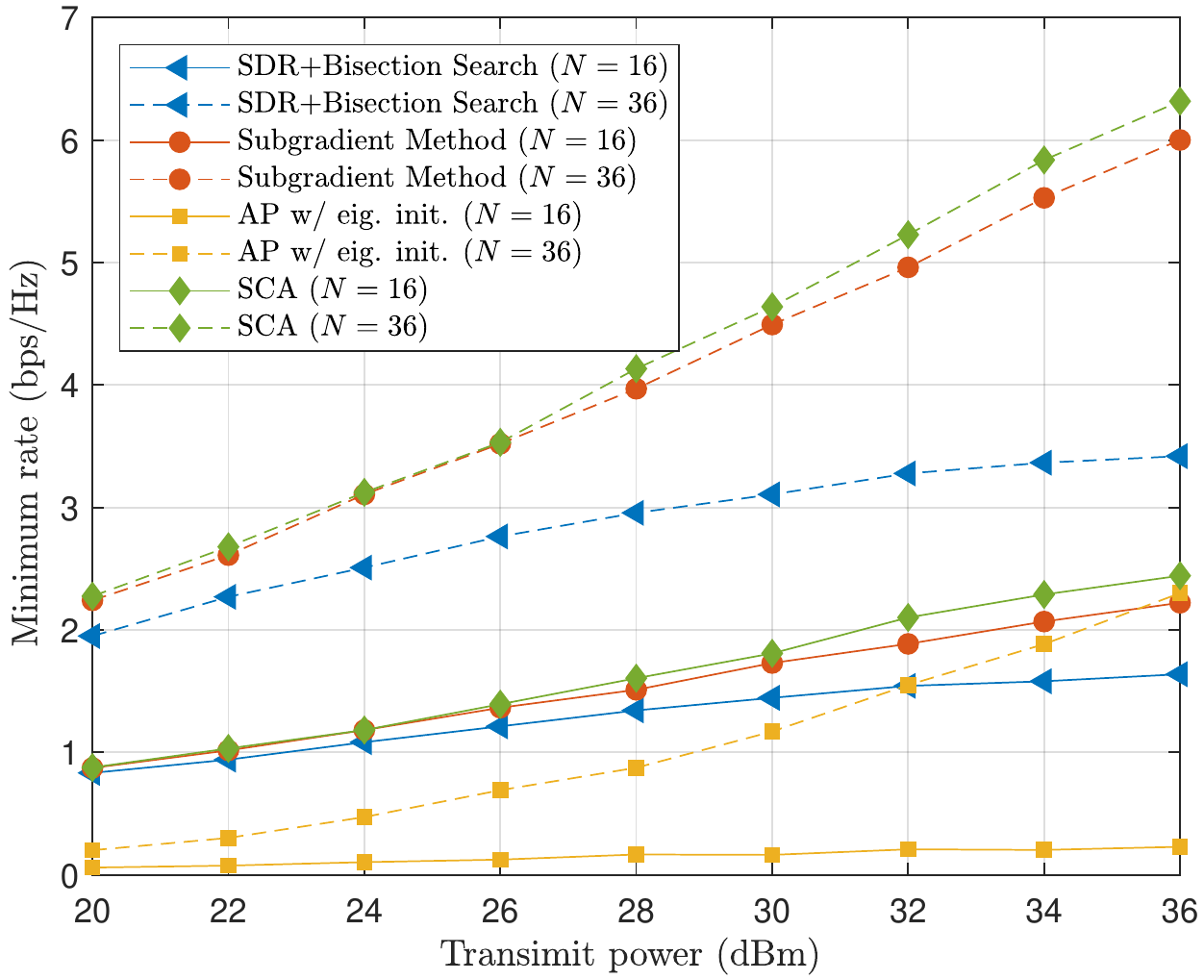}
    	\caption{Minimum rate vs. transmit power in an RIS-enabled interference channel with $K=4$ transceiver pairs without direct links.}
    	\label{fig:minimum_rate}
\end{figure}

\begin{figure*}[t]
    \centering
    \subfigure[Empirical interference nulling probability vs. number of RIS elements~$N$. The number of transceiver pairs $K=7$.\label{fig:phase_transition2_direct}]{\includegraphics[width=8.45cm]{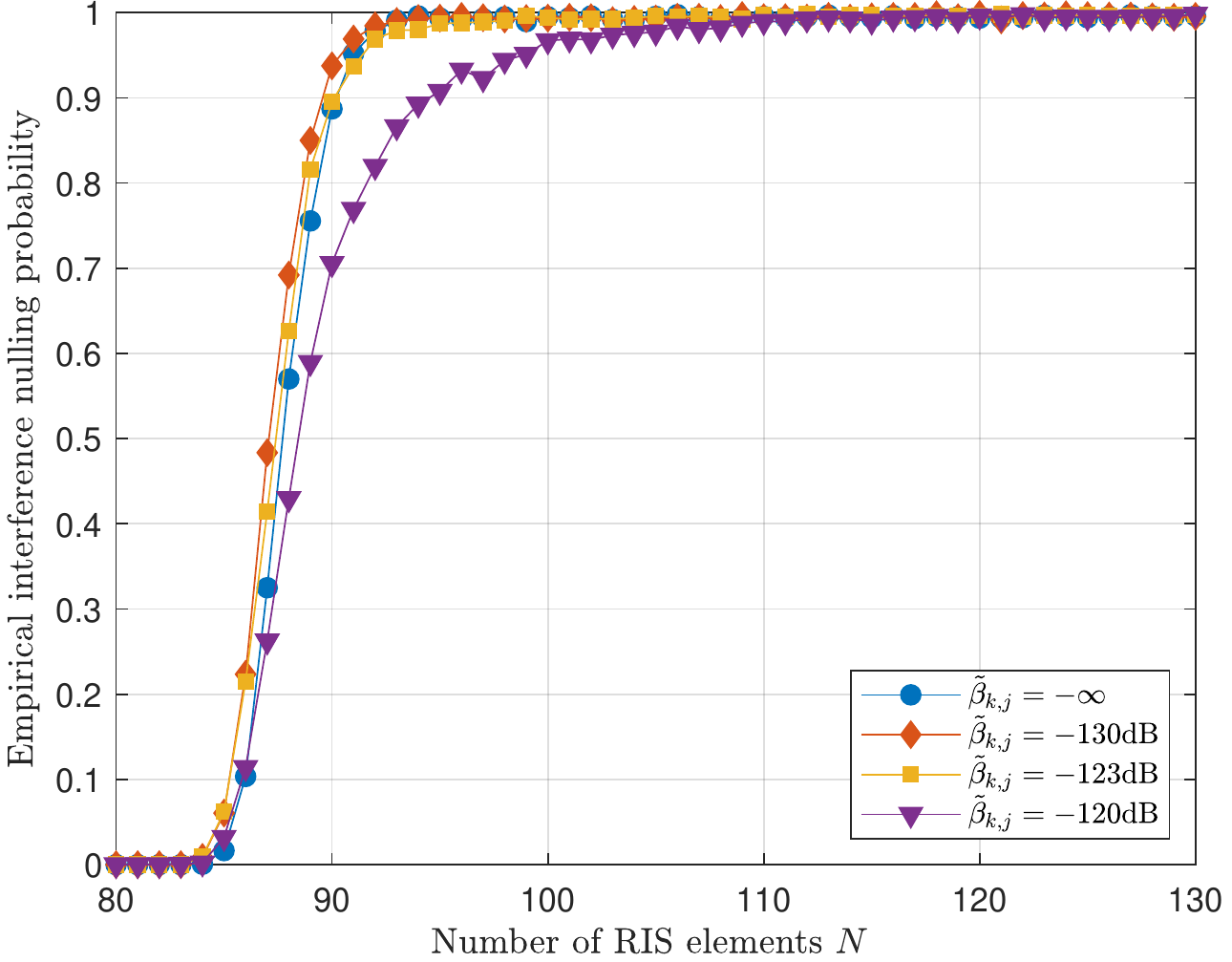}}
    \hspace{2pt}
    \subfigure[Empirical interference nulling probability vs. direct-cascaded link ratio~$\eta$. The number of transceiver pairs $K=8$.  \label{fig:phase_transition1_direct}]{\includegraphics[width=8.6cm]{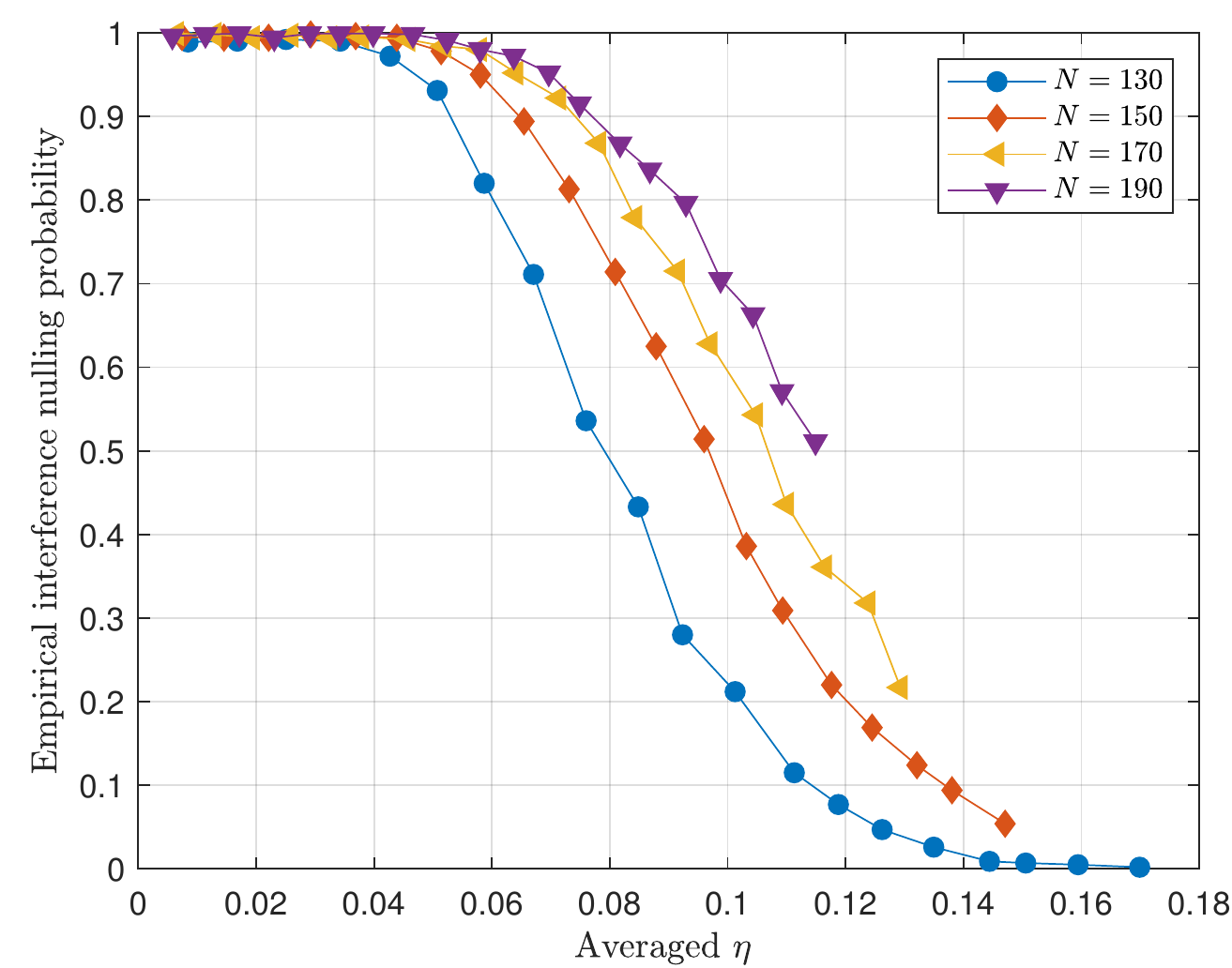}}
    \caption{Phase transition phenomenon for interference nulling in RIS-enabled $K$-user interference channels with direct links.}    
\end{figure*}

\subsection{Minimum-Rate Maximization}\label{subsec:mim_rate_sim}

To verify the effectiveness of the proposed subgradient method for minimum-rate
maximization, we compare its performance with the SDR-based bisection search
method \cite{9246254} \re{and the SCA based approach \cite{yu2020joint}}. 
In the SDR approach, the optimization variable $\bm v$ is lifted to the space
of matrix variables by defining a matrix variable $\bm V=\bm v\bm v^{\sf H}$.
This eliminates the nonconvex unit modulus constraints on $\bm v$.  Semidefinite
programming can then be applied in order to find an optimal $\bm V^\natural$.
If SDR returns a matrix with a rank greater than one, we use Gaussian
randomization (with 1000 samples in this case) to obtain a solution $\bm v$.
\re{In the SCA based algorithm, the
unit modulus constraint of problem \eqref{prob:mim_rate} is enforced by a
penalty term in the objective function \cite{yu2020joint}, then the objective
function is approximated successively by a convex function and the resulting 
optimization problem is solved in each iteration. In the simulations, we use 
the CVX \cite{cvx} software with SDPT3 solver to solve this convex optimization 
problem.} For the proposed subgradient method, we set the maximum number of
iterations to be $10000$ and stop the algorithm if there is no improvement on
the objective function in $1000$ consecutive iterations.

In Fig.~\ref{fig:minimum_rate}, we plot the average minimum rate versus the transmit power. The number of transceiver pairs is set to $K=4$. We consider two scenarios where the number of RIS elements is $N=4\times4$ and $N=6\times 6$. In the former case, as can be observed from Fig.~\ref{fig:phase_transition1},  the interference cannot be completely nulled. In the latter case, we can find a  zero-forcing solution using the alternating projection algorithm with high probability. 

From Fig.~\ref{fig:minimum_rate}, it is clear that zero-forcing by itself does
not produce a good minimum rate. We also observe that both the SCA and the 
subgradient method outperforms the SDR based bisection search method significantly. Furthermore, the performance gain of the SCA and the subgradient method over the SDR-based bisection search method increases as $N$ increases from $16$ to $36$. This is because the SDR approach does not necessarily find a rank-one solution, especially for large values of $N$. \re{At high SNR, the SCA algorithm is slightly better than the subgradient method. However, the SCA method needs to solve a convex optimization problem at each iteration. This leads to an order of magnitude higher complexity than the running time of the subgradient method. Thus, the subgradient method is the best choice from a complexity-performance tradeoff perspective. }

\subsection{RIS System with Transmitter-Receiver Direct Paths}\label{subsec:sim_direct}

\begin{figure}[t]
    \centering
    \includegraphics[width=8.6cm]{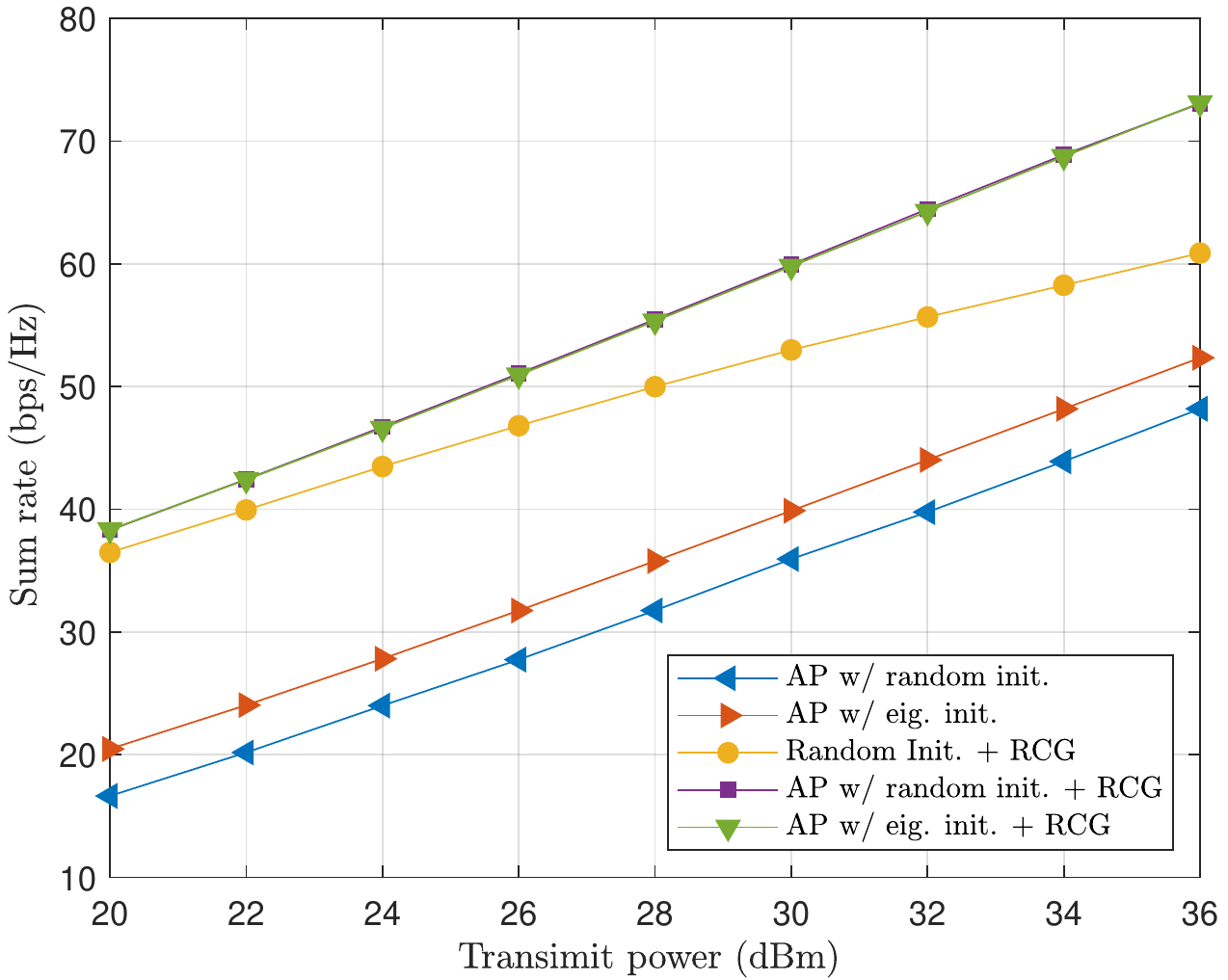}
    \caption{Sum rate vs. transmit power for an RIS system with direct links. The number of RIS elements is $N=12\times 12$ and the number of transceiver pairs is $K=8$.}    
\end{figure}

In this subsection, we evaluate the performance of the proposed algorithms on scenarios with direct paths between the transmitters and the receivers. The direct channels are modeled as Rayleigh fading channels, i.e.,
\re{\begin{equation}
    b_{k,j}= \tilde{\beta}_{k,j}\tilde{b}_{k,j},
\end{equation}
where $\tilde{\beta}_{k,j}$ is the path-loss and $\tilde{b}_{k,j}\sim\mathcal{CN}(0,1)$.}

To see the influence of the strength of the direct channel on the feasibility of zero-forcing solutions, we set the path-loss values of the direct links to be $\tilde{\beta}_{k,j}=-\infty, -130\text{dB}, -123\text{dB}, -120\text{dB}$. In comparison, the average path-loss of the cascaded link is about $-122.6\text{dB}$. The number of transceiver pairs is $K=7$, and again the RIS is assumed to be a linear array for convenience. From Fig.~\ref{fig:phase_transition2_direct}, we can see that the phase transition location does not change if the strength of the direct path is relatively small. But, the phase transition location moves toward higher values of $N$ if the strength of the direct path exceeds some threshold. This implies that more elements at the RIS are needed to null all the interference if the strength of the direct path exceeds some threshold value. 

To characterize this phenomenon more precisely, let $\eta$ denote the maximum ratio between the strength of the direct path and the corresponding $\ell_1$ norm of the cascaded channels, defined as follows:
\begin{align}
    \eta = \max_{k,j} \frac{|b_{k,j}|}{\|\bm a_{k,j}\|_1}.
\end{align} 
In the next experiment, the number of transceiver pairs is $K=8$. By varying the path-loss of the direct paths to be $\tilde{\beta}_{k,j}=(-120+10\log(0.1))\text{dB},(-120+10\log(0.2))\text{dB},\dots,(-120+10\log(2))\text{dB}$, we record the average $\eta$ over $1000$ channel realizations and the corresponding empirical probability of finding a zero-forcing solution. In Fig.~\ref{fig:phase_transition1_direct}, we plot the empirical interference nulling probability versus the average $\eta$. It can be seen that the probability of finding a zero-forcing solution decreases as the value of the average $\eta$ increases. However, we can increase the number of elements at the RIS to compensate. 

Finally, we evaluate the performance of the proposed algorithm for the sum-rate maximization problem in the presence of the direct paths. The number of RIS elements is $N=12\times12$, and the number of transceiver pairs is $K=8$. The path-loss of the direct path is fixed to be $-120\text{dB}$. As can be seen from Fig.~\ref{fig:phase_transition1_direct}, we observe a similar phenomenon as the case without direct paths in Fig.~\ref{fig:sumrate_a}, e.g., RCG with zero-forcing solution as the initialization achieves the best sum-rate performance. This shows that the proposed algorithm can be readily applied to scenarios with direct channels between the transmitters and the receivers.

\section{Conclusion}\label{sect:concllusion}

This paper investigates an RIS-enabled multiple-access environment modeled as a
$K$-user interference channel, where $K$ transceiver pairs communicate through
reflection by the RIS. We show that it is possible to configure the reflective
coefficients of the RIS to completely null interference and to achieve $K$
DoF, if the number of elements at the RIS is sufficiently large.  Specifically,
for the line-of-sight channel model without the direct paths, we show that
there exists a solution to achieve $K$ DoF if the number of RIS elements
exceeds some finite value that depends only on $K$. Further, we propose an
alternating projection algorithm with local convergence guarantee to find the
zero-forcing solution for any arbitrary channel realizations.  Numerical
results demonstrate that the proposed alternating projection algorithm can
achieve interference nulling if the number of RIS elements $N$ is slightly
larger than $2K(K-1)$. 

This paper also proposes an efficient two-stage scheme to maximize the network
sum rate. We show experimentally that using the zero-forcing solution obtained
by the alternating projection algorithm as the initial point for the subsequent
RCG method can significantly improve the optimization performance as compared to
random initialization. For the problem of maximizing the minimum rate, we
propose a subgradient projection method which is scalable to large $N$. In all,
the results of this paper demonstrate the considerable capability of the RIS
for interference nulling in an RIS-assisted multiuser communications environment.

\bibliographystyle{IEEEtran}
\bibliography{ref}

\end{document}